\input harvmac.tex
\newdimen\tableauside\tableauside=1.0ex
\newdimen\tableaurule\tableaurule=0.4pt
\newdimen\tableaustep
\def\phantomhrule#1{\hbox{\vbox to0pt{\hrule height\tableaurule width#1\vss}}}
\def\phantomvrule#1{\vbox{\hbox to0pt{\vrule width\tableaurule height#1\hss}}}
\def\sqr{\vbox{%
  \phantomhrule\tableaustep
  \hbox{\phantomvrule\tableaustep\kern\tableaustep\phantomvrule\tableaustep}%
  \hbox{\vbox{\phantomhrule\tableauside}\kern-\tableaurule}}}
\def\squares#1{\hbox{\count0=#1\noindent\loop\sqr
  \advance\count0 by-1 \ifnum\count0>0\repeat}}
\def\tableau#1{\vcenter{\offinterlineskip
  \tableaustep=\tableauside\advance\tableaustep by-\tableaurule
  \kern\normallineskip\hbox
    {\kern\normallineskip\vbox
      {\gettableau#1 0 }%
     \kern\normallineskip\kern\tableaurule}%
  \kern\normallineskip\kern\tableaurule}}
\def\gettableau#1 {\ifnum#1=0\let\next=\null\else
  \squares{#1}\let\next=\gettableau\fi\next}

\tableauside=1.0ex
\tableaurule=0.4pt
\input epsf
\noblackbox
\def\p{\partial}

\def\ra{{\rightarrow}}

\def\p{\partial}



\def\unlockat{\catcode`\@=11}
\def\lockat{\catcode`\@=12}

\unlockat

\def\newsec#1{\global\advance\secno by1\message{(\the\secno. #1)}
\global\subsecno=0\global\subsubsecno=0\eqnres@t\noindent
{\bf\the\secno. #1}
\writetoca{{\secsym} {#1}}\par\nobreak\medskip\nobreak}
\global\newcount\subsecno \global\subsecno=0
\def\subsec#1{\global\advance\subsecno
by1\message{(\secsym\the\subsecno. #1)}
\ifnum\lastpenalty>9000\else\bigbreak\fi\global\subsubsecno=0
\noindent{\it\secsym\the\subsecno. #1}
\writetoca{\string\quad {\secsym\the\subsecno.} {#1}}
\par\nobreak\medskip\nobreak}
\global\newcount\subsubsecno \global\subsubsecno=0
\def\subsubsec#1{\global\advance\subsubsecno
\message{(\secsym\the\subsecno.\the\subsubsecno. #1)}
\ifnum\lastpenalty>9000\else\bigbreak\fi
\noindent\quad{\secsym\the\subsecno.\the\subsubsecno.}{#1}
\writetoca{\string\qquad{\secsym\the\subsecno.\the\subsubsecno.}{#1}}
\par\nobreak\medskip\nobreak}

\def\subsubseclab#1{\DefWarn#1\xdef
#1{\noexpand\hyperref{}{subsubsection}%
{\secsym\the\subsecno.\the\subsubsecno}%
{\secsym\the\subsecno.\the\subsubsecno}}%
\writedef{#1\leftbracket#1}\wrlabeL{#1=#1}}
\lockat

\def\IL{{\relax{\rm I\kern-.18em L}}}
\def\IH{{\relax{\rm I\kern-.18em H}}}
\def\IR{{\relax{\rm I\kern-.18em R}}}
\def\IE{{\relax{\rm I\kern-.18em E}}}
\def\IC{{\relax\hbox{$\inbar\kern-.3em{\rm C}$}}}
\def\IZ{{\relax\ifmmode\mathchoice
{\hbox{\cmss Z\kern-.4em Z}}{\hbox{\cmss Z\kern-.4em Z}}
{\lower.9pt\hbox{\cmsss Z\kern-.4em Z}}
{\lower1.2pt\hbox{\cmsss Z\kern-.4em Z}}\else{\cmss Z\kern-.4em
Z}\fi}}

\def\CD {{\cal D}}
\def\CF {{\cal F}}

\def\CP {{\cal P }}

\def\CO {{\cal O}}

\def\CC {{\cal C}}

\def\CA{{\cal A}}


\def\CO {{\cal O}}

\def\CP {{\cal P }}

\def\ch{{\rm ch}}

\font\manual=manfnt \def\dbend{\lower3.5pt\hbox{\manual\char127}}

\def\IZ{{\relax\ifmmode\mathchoice
{\hbox{\cmss Z\kern-.4em Z}}{\hbox{\cmss Z\kern-.4em Z}}
{\lower.9pt\hbox{\cmsss Z\kern-.4em Z}}
{\lower1.2pt\hbox{\cmsss Z\kern-.4em Z}}\else{\cmss Z\kern-.4em
Z}\fi}}

\def\p{\partial}

\def\CO {{\cal O}}

\def\CP {{\cal P }}

\def\ch{{\rm ch}}

\def\om{{\overline M}}


\def\IZ{{\relax\ifmmode\mathchoice
{\hbox{\cmss Z\kern-.4em Z}}{\hbox{\cmss Z\kern-.4em Z}}
{\lower.9pt\hbox{\cmsss Z\kern-.4em Z}}
{\lower1.2pt\hbox{\cmsss Z\kern-.4em Z}}\else{\cmss Z\kern-.4em
Z}\fi}}
\def\IB{{\relax{\rm I\kern-.18em B}}}
\def\IC{{\relax\hbox{$\inbar\kern-.3em{\rm C}$}}}
\def\ID{{\relax{\rm I\kern-.18em D}}}
\def\IE{{\relax{\rm I\kern-.18em E}}}
\def\IF{{\relax{\rm I\kern-.18em F}}}
\def\IG{{\relax\hbox{$\inbar\kern-.3em{\rm G}$}}}
\def\IGa{{\relax\hbox{${\rm I}\kern-.18em\Gamma$}}}
\def\IH{{\relax{\rm I\kern-.18em H}}}
\def\II{{\relax{\rm I\kern-.18em I}}}
\def\IK{{\relax{\rm I\kern-.18em K}}}
\def\IP{{\relax{\rm I\kern-.18em P}}}

\def\inbar{\,\vrule height1.5ex width.4pt depth0pt}

\def\p{\partial}

\font\cmss=cmss10 \font\cmsss=cmss10 at 7pt
\def\IR{\relax{\rm I\kern-.18em R}}
\def\IT{\relax{\rm I\kern-.45em T}}


\def\boxit#1{\vbox{\hrule\hbox{\vrule\kern8pt
\vbox{\hbox{\kern8pt}\hbox{\vbox{#1}}\hbox{\kern8pt}}
\kern8pt\vrule}\hrule}}
\def\mathboxit#1{\vbox{\hrule\hbox{\vrule\kern8pt\vbox{\kern8pt
\hbox{$\displaystyle #1$}\kern8pt}\kern8pt\vrule}\hrule}}


\def\inbar{\,\vrule height1.5ex width.4pt depth0pt}

\def\p{\partial}

\font\cmss=cmss10 \font\cmsss=cmss10 at 7pt
\def\IR{\relax{\rm I\kern-.18em R}}

\def\pib{{\bar{\pi}}}


\def\ra{{\longrightarrow}}

\let\includefigures=\iftrue
\newfam\black
\includefigures

\input epsf
\def\plb#1 #2 {Phys. Lett. {\bf B#1} #2 }
\long\def\del#1\enddel{}
\long\def\new#1\endnew{{\bf #1}}
\let\<\langle \let\>\rangle

\def\figin{\epsfcheck\figin}\def\figins{\epsfcheck\figins}
\def\epsfcheck{\ifx\epsfbox\UnDeFiNeD
\message{(NO epsf.tex, FIGURES WILL BE IGNORED)}
\gdef\figin##1{\vskip2in}\gdef\figins##1{\hskip.5in} blank space instead
\else\message{(FIGURES WILL BE INCLUDED)}
\gdef\figin##1{##1}\gdef\figins##1{##1}\fi}
\def\DefWarn#1{}
\def\figinsert{\goodbreak\midinsert}
\def\ifig#1#2#3{\DefWarn#1\xdef#1{fig.~\the\figno}
\writedef{#1\leftbracket fig.\noexpand~\the\figno}
\figinsert\figin{\centerline{#3}}\medskip
\centerline{\vbox{\baselineskip12pt
\advance\hsize by -1truein\noindent
\footnotefont{\bf Fig.~\the\figno:} #2}}
\bigskip\endinsert\global\advance\figno by1}
\else
\def\ifig#1#2#3{\xdef#1{fig.~\the\figno}
\writedef{#1\leftbracket fig.\noexpand~\the\figno}
\figinsert\figin{\centerline{#3}}\medskip
\centerline{\vbox{\baselineskip12pt
\advance\hsize by -1truein\noindent
\footnotefont{\bf Fig.~\the\figno:} #2}}
\bigskip\endinsert
\global\advance\figno by1}
\fi

\input xy
\xyoption{all}
\font\cmss=cmss10 \font\cmsss=cmss10 at 7pt
\def\inbar{\,\vrule height1.5ex width.4pt depth0pt}
\def\IC{{\relax\hbox{$\inbar\kern-.3em{\rm C}$}}}
\def\IP{{\relax{\rm I\kern-.18em P}}}
\def\IF{{\relax{\rm I\kern-.18em F}}}
\def\IZ{\relax\ifmmode\mathchoice
{\hbox{\cmss Z\kern-.4em Z}}{\hbox{\cmss Z\kern-.4em Z}}
{\lower.9pt\hbox{\cmsss Z\kern-.4em Z}}
{\lower1.2pt\hbox{\cmsss Z\kern-.4em Z}}\else{\cmss Z\kern-.4em
Z}\fi}
\def\IR{{\relax{\rm I\kern-.18em R}}}
\def\IQ{\relax\hbox{\kern.25em$\inbar\kern-.3em{\rm Q}$}}

\def\pmb#1{\setbox0=\hbox{#1}%
 \kern-.025em\copy0\kern-\wd0
 \kern.05em\copy0\kern-\wd0
 \kern-.025em\raise.0433em\box0 }
\font\cmss=cmss10
\font\cmsss=cmss10 at 7pt
\def\rlx{\relax\leavevmode}
\def\Cop{\relax\,\hbox{$\inbar\kern-.3em{\rm C}$}}
\def\Rop{\relax{\rm I\kern-.18em R}}
\def\Nop{\relax{\rm I\kern-.18em N}}
\def\Pop{\relax{\rm I\kern-.18em P}}
\def\Zop{\rlx\leavevmode\ifmmode\mathchoice{\hbox{\cmss Z\kern-.4em Z}}
 {\hbox{\cmss Z\kern-.4em Z}}{\lower.9pt\hbox{\cmsss Z\kern-.36em Z}}
 {\lower1.2pt\hbox{\cmsss Z\kern-.36em Z}}\else{\cmss Z\kern-.4em
 Z}\fi}

\def\inbar{\,\vrule height1.5ex width.4pt depth0pt}
\def\IC{{\relax\hbox{$\inbar\kern-.3em{\rm C}$}}}
\def\IP{{\relax{\rm I\kern-.18em P}}}
\def\IF{{\relax{\rm I\kern-.18em F}}}
\def\IZ{\relax\ifmmode\mathchoice
{\hbox{\cmss Z\kern-.4em Z}}{\hbox{\cmss Z\kern-.4em Z}}
{\lower.9pt\hbox{\cmsss Z\kern-.4em Z}}
{\lower1.2pt\hbox{\cmsss Z\kern-.4em Z}}\else{\cmss Z\kern-.4em
Z}\fi}
\def\IR{{\relax{\rm I\kern-.18em R}}}
\def\IT{{\mathchoice {\setbox0=\hbox{$\displaystyle\rm
T$}\hbox{\hbox to0pt{\kern0.3\wd0\vrule height0.9\ht0\hss}\box0}}
{\setbox0=\hbox{$\textstyle\rm T$}\hbox{\hbox
to0pt{\kern0.3\wd0\vrule height0.9\ht0\hss}\box0}}
{\setbox0=\hbox{$\scriptstyle\rm T$}\hbox{\hbox
to0pt{\kern0.3\wd0\vrule height0.9\ht0\hss}\box0}}
{\setbox0=\hbox{$\scriptscriptstyle\rm T$}\hbox{\hbox
to0pt{\kern0.3\wd0\vrule height0.9\ht0\hss}\box0}}}}
\def\bbbti{{\mathchoice {\setbox0=\hbox{$\displaystyle\rm
T$}\hbox{\hbox to0pt{\kern0.3\wd0\vrule height0.9\ht0\hss}\box0}}
{\setbox0=\hbox{$\textstyle\rm T$}\hbox{\hbox
to0pt{\kern0.3\wd0\vrule height0.9\ht0\hss}\box0}}
{\setbox0=\hbox{$\scriptstyle\rm T$}\hbox{\hbox
to0pt{\kern0.3\wd0\vrule height0.9\ht0\hss}\box0}}
{\setbox0=\hbox{$\scriptscriptstyle\rm T$}\hbox{\hbox
to0pt{\kern0.3\wd0\vrule height0.9\ht0\hss}\box0}}}}
\def\K{{\cal{K}}}
\def\R{{\cal{R}}}

\def\M{{\cal M}}
\def\C{{\cal C}}
\def\H{{\cal H}}

\def\op{{\overline P}}
\def\oq{{\overline Q}} 
\def\oa{{\overline A}}
\def\oe{{\overline E}}

\def\ovr{{\overline R}}
 
\def\rra#1{
  \setbox1=\hbox{\kern10pt${#1}$\kern10pt}
  \,\vbox{\offinterlineskip\hbox to\wd1{\hfil\copy1\hfil}
    \kern 3pt\hbox to\wd1{\rightarrowfill}}\,}

\nref\Ai{P.S. Aspinwall, ``A Point's Point of View of Stringy Geometry'', 
JHEP {\bf 01} (2003) 002, hep-th/0203111.}
\nref\Aii{P.S. Aspinwall, ``The Breakdown of Topology at Small Scales'', hep-th/0312188.}
\nref\AD{P. S. Aspinwall and M. R. Douglas, ``D-brane stability and monodromy,'' 
JHEP {\bf 0205}, 031 (2002), hep-th/0110071.}
\nref\AL{P. S. Aspinwall and A. E. Lawrence, ``Derived categories and zero-brane stability,'' 
JHEP {\bf 0108}, 004 (2001), hep-th/0104147.}
\nref\AHK{P.S. Aspinwall, R.P. Horja and R.L. Karp, ``Massless D-Branes on Calabi-Yau Threefolds 
and Monodromy'', hep-th/0209161.}
\nref\B{A.A. Beilinson, ``Coherent Sheaves on $\IP^n$ and Problems in Linear 
Algebra'', Funk. An. {\bf 12} (1978) 68.} 
\nref\Bo{A. I. Bondal, ``Helixes, Representations of Quivers and Koszul Algebras'', 
{\it Felices and Vector Bundles}, A.N. Rudakov ed, London Mathematical Society 
Lecture Series {\bf 148} 1990.}
\nref\BKR{T. Bridgeland, A. King and M. Reid, 
``Mukai implies McKay: the McKay correspondence as an equivalence of derived categories'', 
J. Amer. Math. Soc. {\bf 14} (2001) 535, math.AG/9908027.}
\nref\BDi{I. Brunner and J. Distler, 
``Torsion D-Branes in Nongeometrical Phases'', 
Adv. Theor. Math. Phys. {\bf 5} (2002) 265, hep-th/0102018.}
\nref\BDii{I. Brunner, J. Distler and R. Mahajan, ``Return of the Torsion D-Branes'', 
Adv. Theor. Math. Phys. {\bf 5} (2002) 311, hep-th/0106262.}
\nref\BDLR{I. Brunner, M. R. Douglas, A. E. Lawrence and C. Romelsberger,
``D-branes on the quintic,'' JHEP {\bf 0008}, 015 (2000), hep-th/9906200.}
\nref\BHLS{I. Brunner, M. Herbst, W. Lerche and B. Scheuner, ``Landau-Ginzburg
Realization of Open String TFT'', hep-th/0305133.}
\nref\BSi{I. Brunner and V. Schomerus,
``D-branes at Singular Curves of Calabi-Yau Compactifications'', 
JHEP {\bf 04} (2000) 020, hep-th/0001132.}
\nref\BSii{I. Brunner and V. Schomerus, 
``On Superpotentials for D-Branes in Gepner Models'', JHEP {\bf 10} (2000) 016,hep-th/0008194}
\nref\ESiv{A. Caldararu, S. Katz and E. Sharpe, ``D-branes, B fields, and Ext groups'', 
Adv. Theor. Math. Phys. {\bf 7} (2003) 381}%
\nref\DED{D.-E. Diaconescu, ``Enhanced D-brane Categories from String Field Theory'', 
JHEP {\bf 06} (2001) 016, hep-th/0104200.}
\nref\DD{D.-E. Diaconescu and M. Douglas, 
``D-Branes on Stringy Calabi-Yau Manifolds'', hep-th/0006224.} 
\nref\DG{D.-E. Diaconescu, J. Gomis, 
``Fractional Branes and Boundary States in Orbifold Theories'', 
hep-th/9906242.}%
\nref\DR{D.-E. Diaconescu, C. Romelsberger, 
``D-Branes and Bundles on Elliptic Fibrations'', 
Nucl. Phys. {\bf B574} (2000) 245, hep-th/9910172.}%
\nref\DJP{J. Distler, H. Jockers and H. Park, ``D-Brane Monodromies, Derived Categories and 
Boundary Linear Sigma Models'', hep-th/0206242.} 
\nref\ESv{R. Donagi, S. Katz, E. Sharpe, ``Spectra of D-branes with Higgs vevs'', 
hep-th/0309270}%
\nref\D1{M. R. Douglas, ``D-branes, categories and N = 1 supersymmetry,'' 
J.\ Math.\ Phys.\  {\bf 42}, 2818 (2001), hep-th/0011017.}
\nref\DGJT{M. R. Douglas, S. Govindarajan, T. Jayaraman and A. Tomasiello, 
``D-branes on Calabi-Yau manifolds and superpotentials,'', hep-th/0203173.}
\nref\DF{M. R. Douglas and B. Fiol, `` D-branes and Discrete Torsion II'', hep-th/9903031.}
\nref\DFRi{M. R. Douglas, B. Fiol, C. Romelsberger, 
``Stability and BPS branes'', hep-th/0002037.}%
\nref\DFRii{M. R. Douglas, B. Fiol, C. Romelsberger,
``The spectrum of BPS branes on a noncompact Calabi-Yau'', 
hep-th/0003263.}%
\nref\DM{M.R. Douglas and G. Moore, ``D-branes, Quivers, and ALE Instantons'', hep-th/9603167.}%
\nref\GJcycles{S. Govindarajan, T. Jayaraman, Tapobrata Sarkar, 
``Worldsheet approaches to D-branes on supersymmetric cycles'', 
Nucl. Phys. {\bf B580} (2000) 519,hep-th/9907131} 
\nref\GJi{S. Govindarajan and T. Jayaraman, ``Boundary fermions, coherent sheaves and D-branes on 
Calabi-Yau  manifolds,'' Nucl.\ Phys.\ B {\bf 618}, 50 (2001), hep-th/0104126.}
\nref\GJ{S. Govindarajan, T. Jayaraman, ``On the Landau-Ginzburg 
description of Boundary CFTs and special Lagrangian submanifolds'', 
JHEP {\bf 07} (2000) 016, hep-th/0003242;
``D-branes, Exceptional Sheaves and Quivers on Calabi-Yau manifolds: 
From Mukai to McKay'', hep-th/0010196; ``Boundary Fermions, 
Coherent Sheaves and D-branes on Calabi-Yau manifold'', hep-th/0104126.}%
\nref\GJS{S. Govindarajan, T. Jayaraman, and T. Sarkar, 
``On D-branes from Gauged Linear Sigma Models'', Nucl. Phys. {\bf B593} 
(2001) 155, hep-th/0007075.}%
\nref\GL{B. Greene, C. Lazaroiu, ``Collapsing D-Branes in 
Calabi-Yau Moduli Space: I'', hep-th/0001025.}%
\nref\GS{M. Gutperle, Y. Satoh, ``D-branes in Gepner models and 
supersymmetry'', Nucl. Phys. {\bf B543} (1999) 73
hep-th/9808080.}
\nref\HM{S. Hellerman and J. McGreevy, ``Linear sigma model toolshed for D-brane physics'', 
JHEP {\bf 10} (2001) 002, hep-th/0104100.}%
\nref\HKLM{S. Hellerman, S. Kachru, A. Lawrence and J. McGreevy,
''Linear Sigma Models for Open Strings'',
JHEP {\bf 07} (2002) 002, hep-th/0109069.}
\nref\H{K. Hori, ``Linear Models of Supersymmetric D-Branes'', hep-th/0012179.}
\nref\HV{K. Hori and C. Vafa, ``Mirror Symmetry'', hep-th/0002222.}
\nref\HIV{K. Hori, A. Iqbal and C. Vafa, ``D-branes and Mirror Symmetry'', hep-th/0005247.}
\nref\RPHi{R.P. Horja, ``Hypergeometric functions and mirror 
symmetry in toric varieties'', math.AG/9912109.}
\nref\RPHii{R.P. Horja,  
``Derived Category Automorphisms from Mirror Symmetry'', 
math.AG/0103231.}
\nref\KLI{A. Kapustin and Y. Li, ``D-Branes in Landau-Ginzburg Models and Algebraic Geometry'', 
hep-th/0210296.}
\nref\KLII{A. Kapustin and Y. Li, 
``Topological Correlators in Landau-Ginzburg Models with Boundaries'', 
hep-th/035136.}
\nref\KLIII{A. Kapustin and Y. Li, 
``D-Branes in Topological Minimal Models: The Landau-Ginzburg Approach'', 
hep-th/0306001.}
\nref\KLLW{P. Kaste, W. Lerche, C.A. Lutken, J. Walcher, 
``D-Branes on K3-Fibrations'', Nucl. Phys. {\bf B582} (2000) 
203, hep-th/9912147.}%
\nref\ESii{S. Katz and E. Sharpe, ``D-branes, open string vertex operators, and Ext groups'', 
Adv. Theor. Math. Phys. {\bf 6} (2003) 979, hep-th/0208104}%
\nref\ESiii{S. Katz, T. Pantev and E. Sharpe, ``D-branes, orbifolds, and Ext groups'', 
Nucl. Phys. {\bf B673} (2003) 263, hep-th/0212218}%
\nref\K{M. Kontsevich, ``Homological Algebra of Mirror Symmetry'', 
alg-geom/9411018.}
\nref\CLv{C. Lazaroiu, 
``Collapsing D-branes in one-parameter models and small/large 
radius duality'',  hep-th/0002004.}%
\nref\CLi{C.I. Lazaroiu, ``Unitarity, D-brane dynamics and D-brane categories'', 
JHEP {\bf 12} (2001) 031, hep-th/0102183.}
\nref\CLii{C.I. Lazaroiu, ``Generalized complexes and string field theory'', 
JHEP {\bf 06} (2001) 052, hep-th/0102122.}
\nref\M{P. Mayr, ``Phases of Supersymmetric D-branes on Kaehler 
Manifolds and the McKay correspondence'', JHEP {\bf 01} (2001) 018,
hep-th/0010223.}%
\nref\O{D. Orlov, ``Triangulated Categories of Singularities and 
D-Branes in Landau-Ginzburg Orbifolds'', math.AG/0302304.}
\nref\R{A. Recknagel, ``Permutation Branes'', JHEP {\bf 04} (2003) 041, hep-th/0208119.}%
\nref\RS{A. Recknagel and V. Schomerus, ``D-branes in Gepner models,'' 
Nucl.\ Phys.\ B {\bf 531}, 185 (1998), 
hep-th/9712186.}
\nref\ESa{E. Scheidegger, ``D-branes on some 
one- and two-parameter Calabi-Yau hypersurfaces'', 
JHEP {\bf 04} (2000) 003, 
hep-th/9912188.}
\nref\ESb{E. Scheidegger, ``On D0-branes in Gepner models'', JHEP {\bf 08} (2002) 001, 
hep-th/0109013.}%
\nref\ESi{E. Sharpe, ``D-Branes, Derived Categories, and Grothendieck Groups'', 
Nucl. Phys. {\bf B561} (1999) 433, hep-th/9902116}%
\nref\ST{P. Seidel and R. Thomas, 
``Braid Group Actions on Derived Categories of Coherent Sheaves'', 
math.AG/0001043.}
\nref\T{A. Tomasiello, ``D-branes on Calabi-Yau manifolds and helices'', 
JHEP {\bf 02} (2001) 008, hep-th/0010217.}
\nref\W1{E. Witten, ``Phases of N = 2 theories in two dimensions,'' 
Nucl.\ Phys.\ B {\bf 403}, 159 (1993), 
hep-th/9301042.}
\Title{
\vbox{
\baselineskip12pt
\hbox{hep-th/0401135}
\hbox{RUNHETC-2004-02}}}
{\vbox{\vskip 37pt
\vbox{\centerline{Fractional Branes in Landau-Ginzburg Orbifolds}}
}}
\vskip 15pt
\centerline{Sujay K. Ashok, Eleonora Dell'Aquila and 
Duiliu-Emanuel Diaconescu\footnote{$^{}$}{{{\tt ashok,dellaqui,duiliu@physics.rutgers.edu}}}}
\bigskip
\medskip
\centerline{{\it Department of Physics and Astronomy,
Rutgers University,}}
\centerline{\it Piscataway, NJ 08855-0849, USA}
\bigskip
\bigskip
\bigskip
\bigskip
\smallskip
\noindent 
We construct fractional branes in Landau-Ginzburg orbifold categories and study their 
behavior under marginal closed string perturbations. This approach is shown to be more general 
than the rational boundary state construction. In particular we find new D-branes 
on the quintic -- such as a single D0-brane -- 
which are not restrictions of bundles on the ambient projective space.  
We also exhibit a family of deformations of the D0-brane in the Landau-Ginzburg 
category parameterized by points on the Fermat quintic. 

\vfill
\Date{January 2004}

\newsec{Introduction}

It is by now well established that D-branes in topological string theories form 
a triangulated category \refs{\K,\D} (see also \refs{\ESi,\AL,\CLii,\CLi,\DED,\ESii,\ESiii,\ESiv,\ESv}.) 
This algebraic structure captures very important aspects 
of D-brane dynamics such as brane/anti-brane annihilation and bound state formation. 
For topological ${\bf B}$-models on Calabi-Yau threefolds, it has been shown that 
the D-brane category is the bounded derived category of the target manifold 
\refs{\K,\D,\AL,\ESi}.

It is also well known that Calabi-Yau compactifications are continuously connected 
to Landau-Ginzburg orbifolds (also called nongeometric phases) 
by marginal closed string perturbations \W.\ A central problem 
in this context is concerned with the behavior of D-branes under such perturbations. 
Answering this question requires a good control over D-brane 
dynamics in nongeometric phases. One possible approach to this problem relies on  
boundary states in Gepner models \refs{\RS,\GS} and quiver gauge theories \DM.\
The main idea is that one can represent all rational boundary states as 
composites of a finite collection of 
elementary branes, called fractional branes. Many aspects of bound state formation 
are remarkably captured by quiver gauge theory dynamics. These methods have been 
successfully applied to D-branes on Calabi-Yau hypersurfaces in a series of papers 
\refs{\BDi-\BSii, \DED-\GL,\CLv,\M,\ESa}. 
One of the main outcomes of \refs{\BDLR,\DD,\DFRii,\DG,\DD,\M,\GJ,\T} 
is that fractional branes are related by analytic 
continuation to an exceptional collection of bundles 
on the ambient weighted projective space. From a mathematical point of view, this can be understood 
as a derived McKay correspondence \refs{\BKR,\ESiii}. 

An alternative approach to D-branes in topological Landau-Ginzburg models has been recently 
developed in \refs{\KLI-\KLIII,\O}. 
The main result is that topological Landau-Ginzburg D-branes form 
a category which admits an abstract algebraic description based on the Landau-Ginzburg 
superpotential. This has been shown to be a very effective approach to D-branes 
in minimal models \refs{\KLIII}. 

In this paper we consider D-brane categories associated to Landau-Ginzburg 
orbifolds defined by quasihomogeneous superpotentials. Such models are typically 
encountered in the context of gauged linear sigma models.
We begin with a discussion of Landau-Ginzburg boundary conditions and D-brane 
categories in section two.  
In section three we give an explicit 
algebraic construction of fractional branes which can be easily extended to more general 
Gepner model rational boundary states. 

In section four we construct more general objects in Landau-Ginzburg orbifold categories
-- called new fractional branes -- 
which do not have a rational boundary state counterpart. In order to clarify 
their role in the theory, in section five we determine their geometric interpretation 
in the large radius limit using topological and algebraic techniques. The most important
point is that such objects correspond to bundles (or more general derived objects) 
which are not restrictions from the ambient toric variety. 
In particular we find that one of these objects corresponds to a 
single D0-brane on the Fermat quintic. This is an important result
since the D0-brane on the quintic cannot be given a rational boundary state construction 
at the Landau-Ginzburg point. However, rational boundary states with the quantum numbers 
of one D0-brane are known to exist in other models. A systematic treatment can be found 
in \refs{\ESb}. Also, boundary states with the quantum numbers of five D0-branes on the 
quintic have been recently constructed in \refs{\R}.  

The algebraic constructions developed here can also be very effectively applied 
to questions regarding deformations and moduli of D-branes in nongeometric phases. 
In order to 
illustrate some of the main ideas, we discuss two such applications in section six. 
First we prove a conjecture of 
\refs{\DGJT} regarding composites of fractional branes. Then we show that the  
Landau-Ginzburg D0-brane admits a family of deformations 
parameterized by points on the Fermat quintic. This is 
a remarkable confirmation of the constructions employed in this paper. It also suggests  
that the Landau-Ginzburg D0-brane may be the appropriate notion of point \refs{\Ai} 
in nongeometric phases. It would be very interesting to explore this idea in more depth 
in connection with \refs{\Ai,\Aii}. 

{\it Acknowledgments.} We would like to thank 
Bogdan Florea, Paul Horja, Emiliano Imeroni, Anton Kapustin, Ludmil Katzarkov, 
John McGreevy, Sameer Murthy, Greg Moore, Tony Pantev and especially 
Mike Douglas for very useful conversations and suggestions. 
The work of D.-E.D. is partially supported by DOE grant DE-FG02-96ER40949 
and an Alfred P. Sloan foundation fellowship. 
D.-E.D. would also like to acknowledge the hospitality 
of KITP Santa Barbara where part of this work was performed. 

\newsec{D-branes categories in Landau-Ginzburg Models}


The starting point of our discussion is a brief review of 
supersymmetric ${\bf B}$-type boundary states in Landau Ginzburg models 
following \refs{\H,\KLI,\BHLS}. We then present
the construction of Landau-Ginzburg D-brane categories following \refs{\KLI,\O}
and extend it to orbifolds. Orbifold categories have been briefly discussed in \refs{\KLII}, 
but here we need a more systematic treatment. 

Consider a Landau-Ginzburg model with $n+1$ chiral superfields $X=(X_a)$, $a=0,\ldots,n$ 
subject to a polynomial superpotential $W(X)$. We assume that $W$ has only one isolated, 
possibly degenerate, critical point at the origin. 
We would like to formulate this theory on the infinite 
strip $x^0\in [-\infty,\infty]$, $x^1\in [0,\pi]$ so that the full bulk-boundary action 
preserves ${\bf B}$-type supersymmetry with supercharge $Q=\overline{Q}_++\overline{Q}_-$. 
In addition to the standard bulk action
\eqn\sbulk{\eqalign{
S_{bulk} &= \int d^2x\,d^4\theta\,\sum_{a=0}^n\,\overline{X_a}X_a + \int d^2x\,d^2\theta\, W(X_a) \cr
&= \int_{\Sigma} d^2x\,\sum_{a=0}^{n}\left( -\p^{\mu}\overline{X_a}\p_{\mu}X_a + {i\over 2}\bar{\psi}_{-a}({\buildrel \leftrightarrow \over \p_0}+{\buildrel \leftrightarrow \over \p_1})\psi_{-a}+ {i\over 2}\bar{\psi}_{+a}({\buildrel \leftrightarrow \over \p_0}-{\buildrel \leftrightarrow \over \p_1})\psi_{+a}-{1\over 4}|\p_a W|^2 \right) \cr
&\quad\quad\quad\quad\quad\quad\quad -\sum_{a,b=0}^{n} \left({1\over 2} \p_{a}\p_b W\,\psi_{+a}\psi_{-b}+{1\over 2} \p_{a}\p_b \overline{W}\,\bar{\psi}_{+a}\bar{\psi}_{-b}\right)
}}
(following the conventions of \BHLS), supersymmetry constraints require an extra boundary term 
containing some number of boundary fermionic superfields $\Pi_\alpha$, $\alpha=1,\ldots,s$. 
These are nonchiral, i.e. $D\Pi_\alpha=G_{\alpha}(X)\,$, where $G$ is a polynomial function of 
the superfields $X=(X_a)$ restricted to the boundary. 
The boundary action is of the form 
\eqn\bdryactA{
S_{bdry} =
{i\over 4} 
\int_{\p \Sigma} dx^0 \sum_{a=0}^n
\left[\bar{\theta}_a\eta_a - \bar{\eta}_a\theta_a\right]_0^{\pi}+S_\Pi\,,}
where $\eta_a=\psi_{-a}+\,\psi_{+a}\,$,  $\theta_a=\psi_{-a}-\,\psi_{+a}$ and 
\eqn\bdryactB{\eqalign{
S_\Pi &= -{1\over 2}\int_{\p\Sigma} dx^0 d^2\theta\ \sum_{\alpha=1}^s
\overline{\Pi}_\alpha\,\Pi_\alpha -{i\over 2}
\int dx^0 d\theta\ \Pi_\alpha\,F_{\alpha}(X) +\ c.c.\ .\cr}}
The $F_\alpha(X)$ are polynomial boundary interactions. 
It was shown in \refs{\KLI,\BHLS} that the full action 
$S=S_{bulk}+S_{bdry}$ preserves $B$-type supersymmetry if $F_\alpha, G_\alpha$ satisfy the constraint 
\eqn\susyconstr{W = \sum_{\alpha=1}^s F_\alpha G_\alpha + \hbox{const}.}
If $W$ has a single isolated critical point at the origin, 
the constant in the right hand side of \susyconstr\ can be taken zero without loss 
of generality. Therefore ${\bf B}$-branes will be classified by systems of polynomials 
$(F_\alpha, G_\alpha)$ so that $W=\sum_{\alpha=1}^s F_\alpha G_\alpha$.  
Physically, such a brane is realized as the end product of 
open string tachyon condensation on a brane-antibrane pair of rank $r=2^s$.
The $F_\alpha$  describe the
tachyonic profile on the brane world volume. The Chan-Paton factors associated to 
the brane-antibrane pair are realized as the irreducible representation of the complex Clifford algebra
\eqn\anticom{\eqalign{
\{\pi_\alpha,\pi_\beta\} = \{\pib_\alpha,\pib_\beta\} &= 0 \cr
\{\pi_\alpha ,\pib_\beta\} &= \delta_{\alpha\beta}.\cr}}
The boundary contribution to the supercharge is 
\eqn\bdrycharge{
D =\sum_{\alpha=1}^s (\pi_\alpha F_\alpha(X) + \pib_\alpha G_\alpha(X)).}

More generally we can consider a string stretched between two branes specified 
by boundary couplings $(F_\alpha^{(1)}, G_\alpha^{(1)})$, $(F_\alpha^{(2)},G_\alpha^{(2)})$.  
The boundary action for the fermionic superfields is then
\eqn\bdryB{\eqalign{
S_{\Pi} =-{1\over 2}\int_{\p\Sigma} dx^0 d^2\theta\ \sum_{\alpha=1}^s 
\overline{\Pi}_\alpha\,\Pi_\alpha -{i\over 2}
\bigg[& \int_{x^1=\pi} dx^0 d\theta\ \Pi_\alpha\,F_\alpha^{(1)}(X)-\cr
& \int_{x^1=0} dx^0 d\theta\ 
\Pi_\alpha\,F^{(2)}_\alpha(X)\bigg] +\ c.c.\cr}}
We study the spectrum of Ramond ground states in this sector. There is a 
one-to-one correspondence between these states and the physical operators 
in the twisted $B$-model. The BRST operator is 
\eqn\brstA{
Q_{tot}=Q|_{\p\Sigma}+D.}
where the first term is the restriction to the boundary of the bulk supercharge $Q$ 
and the second represents the contribution of the boundary fields.
The physical operators are classified by cohomology classes of $Q_{tot}$ acting 
on off-shell open string states. Since
$Q_{tot}\cdot X=0$, any element of the boundary chiral ring can be expanded as a linear combination of monomials $\pi^I{\bar\pi}^J=\prod_{\alpha=1}^s\pi_\alpha^{I(\alpha)}
{\bar\pi}_\alpha^{J(\alpha)}$, where $I(\alpha), J(\alpha)$ take values $0,1$, 
with coefficients in $\IC[X_a]\,$:
\eqn\morphism{
\Phi= \sum_{I,J}f_{I,J}(X_a)\pi^I{\bar\pi}^J.} 
There is a natural $\IZ/2$ grading on the space of boundary fields defined by 
$\hbox{deg}(\Phi)=\sum_{\alpha=1}^s(I(\alpha)-J(\alpha))$ mod 2. 
Homogeneous elements of degree zero will be called bosonic, or even, 
while homogeneous elements of degree one will be called fermionic, or odd. 
The action of $\CD$ on homogeneous elements is given by 
\eqn\brstB{
D\cdot \Phi= D^{(1)}\cdot\Phi-(-1)^{\hbox{deg}(\Phi)} \Phi\cdot D^{(2)}}
where $D^{(1)},D^{(2)}$ are the boundary BRST operators associated to the two D-branes. 
Using this formula, one can find explicit representatives for BRST cohomology, as discussed 
later in several examples. 

\subsec{D-Brane Categories}

We have so far considered a particular class of boundary conditions associated to tachyon 
condensation on brane/anti-brane pairs of equal rank $r=2^s$. This is a restricted set of 
supersymmetric boundary conditions which can be described in terms of additional 
boundary fermionic superfields. One can obtain more general D-branes as end 
products of tachyon condensation on brane/anti-brane pairs of arbitrary rank. 
Taking into account all such boundary conditions, we obtain a triangulated additive 
category $\CC_W$ which admits the following presentation \refs{\KLI,\O}. 

The objects of $\CC_W$ are given by matrix factorizations of $W$, that is 
pairs $\xymatrix{P_1 \ar@<1ex>[r]^{p_1}& P_0\ar@<1ex>[l]^{p_0}\\}$ 
of free $\IC[X_0,\ldots,X_{N}]$-modules 
so that $p_0p_1=p_1p_0=W$. 
Following \refs{\O} we will denote this data by ${\overline P}$. The massless open string states 
between two D-branes $\op,\oq$, form a $\IZ/2$ graded complex 
\eqn\lgA{
\IH(\op,\oq)=\hbox{Hom}(P_1\oplus P_0,  Q_1\oplus Q_0)=\bigoplus_{i,j=0,1} \hbox{Hom}(P_i,Q_j)} 
where the grading is given by $(i-j)$ mod 2. This complex is equipped with an odd differential 
$D$ which represents the BRST operator of the boundary topological field 
theory. The action of $D$ on a homogeneous element $\Phi$ of degree $k$ 
is given by 
\eqn\lgB{ 
D\cdot\Phi = q\cdot\Phi -(-1)^k \Phi\cdot p}
where $p=p_1\oplus p_0:P_1\oplus P_0\ra P_1\oplus P_0$, $q=q_1\oplus q_0 : Q_1\oplus Q_0 \ra 
Q_1\oplus Q_0$. 
This data defines a DG-category $\CP_W$ \refs{\KLI,\O}. The D-brane category $\CC_W$ is the 
category associated to $\CP_W$ by taking the space of morphisms between two 
objects $(\op,\oq)$ to 
be the degree zero cohomology $H^0(\IH(\op,\oq))$ of the complex \lgA.\
We will use the shorthand notation $H^i(\op,\oq)$, $i=0,1$ for the cohomology groups. 
One can show that $\CC_W$ is an additive triangulated category \refs{\KLI,\O}.
Note that there is an obvious similarity between this formal construction and the 
more physical approach explained in the previous subsection. In order to exhibit 
the matrix factorization associated to a boundary condition of the form \bdryactB,\
let us choose the standard $r=2^s$ dimensional representation of the Clifford algebra 
\anticom.\ Then we can explicitly write the boundary supercharge as a $r\times r$ 
matrix which squares to $W$. 
One of the advantages of the algebraic 
approach is that the rank of brane/anti-brane pairs is not restricted to powers 
of $2$.
Let us record the most important properties of $\CC_W$ for applications to physics.

$i)$ $\CC_W$ is equipped with a shift functor 
$\op\ra \op[1]$ \eqn\shift{\op[1]=\left(
\xymatrix{P_0 \ar@<1ex>[r]^{-p_0}& P_1\ar@<1ex>[l]^{-p_1}\\}\right).}
This is an autoequivalence of the category which maps branes to anti-branes. 

$ii)$ Every morphism $\op{\buildrel \phi\over \ra} \oq$ in  
$\CC_W$ can be completed to a distinguished triangle of the form 
\eqn\triangleA{ 
\op{\buildrel \phi\over \ra} \oq \ra \ovr \ra \op[1]} 
where $\ovr$ is a isomorphic to the cone $C(\phi)$ of $\phi$. 
The cone of a morphism $\op{\buildrel \phi\over \ra} \oq$ is defined 
by 
\eqn\triangleB{ 
\ovr=\left(\xymatrix{Q_1\oplus P_0 \ar@<1ex>[r]^{r_1}& Q_0\oplus P_1 \ar@<1ex>[l]^{r_0}\\}
\right), 
\quad r_1=\left[\matrix{q_1 & \phi_0 \cr 0 & -p_0\cr}\right]\quad 
r_0=\left[\matrix{ q_0 & \phi_1 \cr 0 & -p_1\cr}\right].} 
In physical terms, distinguished triangles describe bound state formation \D.\
More precisely, the existence of a triangle \triangleA\ implies that 
any two objects involved in the construction can form the third by tachyon condensation. 
In particular the K-theory charges of the three objects add to zero. 
In order to decide if a particular condensation process actually takes place or not
we need more data which takes the form of a stability condition \refs{\DFRi,\D,\AD}. 
We will not review this aspect in detail here.  

For further reference, note that the Euler character
\eqn\windex{ 
\chi(\op,\oq) = \hbox{dim}H^0(\op,\oq) -\hbox{dim}H^1(\op,\oq).}
defines an (asymmetric) intersection pairing on objects. 
Physically, this is the Witten index of the open string 
Ramond sector defined by the two branes. 

The above construction can be reformulated in terms of $\IZ/2$ graded modules as follows. 
A $\IZ/2$ graded $\IC[X]$-module $P=(P_1,P_0)$ can be thought of as an ordinary module 
$P=P_1\oplus P_0$ equipped with a $\IC$-linear involution $\eta:P\ra P$, $\eta^2=1$.
The homogeneous parts $P_1,P_0$ are the eigenspaces of $\eta$ corresponding to 
the eigenvalues $+1$ and $-1$ respectively. A pair $\op$ can be similarly thought of as a  
triple $(P,\eta_P,p)$ where $p:P\ra P$ is a $\IC[X]$-module homomorphism satisfying 
\eqn\gradedA{ 
p\,\eta_P+\eta_P\, p=0,\qquad p^2=W(X).}
The $\IZ/2$ graded complex \lgA\ can be similarly regarded 
as the $\IC[X]$-module $\hbox{Hom}(P,Q)$ equipped with an endomorphism $D$ 
and an involution $\eta_{PQ}$ satisfying 
\eqn\gradedB{ 
\eta_{PQ} D + D\eta_{PQ} =0,\qquad  D^2=0.} 
The involution is induced by $\eta_P, \eta_Q$. 
We will find this point of view very useful later in the paper.

\subsec{Orbifold Categories} 

We are interested in Landau-Ginzburg orbifolds obtained by gauging a discrete 
symmetry group $G$ of $W$. Typically, these models are realized as infrared effective 
theories of gauged linear sigma models, in which case $G$ is a finite cyclic group $G=\IZ_d$
for some $d>0$. The construction of the D-brane category can be easily extended to this 
situation. \KLIII.\ The objects are 
pairs $\op$ of $G$-equivariant free $\IC[X]$-modules subject to an equivariant 
condition on the maps $p_0, p_1$. More concretely, regarding $P_0, P_1$ as trivial bundles 
$P_i=\IC^{N+1} \times \IC^{r_i}$, $i=0,1$ of rank $r_0, r_1$, we have to specify representations 
$R_i$ of $G$ on $\IC^{r_i}$, $i=0,1$. We can represent the object in the orbifold category as 

\ifig\OrbifoldObj{Orbifolded Object}{\epsfxsize2.0in\epsfbox{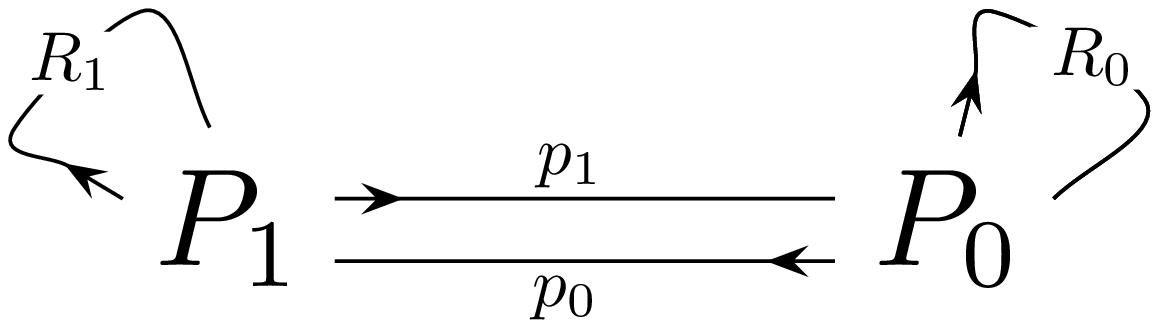}}

\noindent
If we denote by $\rho:G\ra GL(n+1,\IC)$ 
the representation of $G$ on $\IC^{n+1}$, the maps $p_0, p_1$ must satisfy the 
equivariance condition
\eqn\lgC{ 
R_1(g)p_0(\rho(g^{-1})X) R_0(g^{-1})=p_0(X), \qquad 
R_0(g)p_1(\rho(g^{-1})X) R_1(g^{-1})=p_1(X)} 
for any group element $g\in G$. 
This condition imposes certain restrictions on the allowed representations 
$R_0, R_1$, as explained later in examples. 
Given two such objects $\op,\oq$, the action of $G$ on $P_i, Q_i$, $i=0,1$ 
induces an action on 
the terms in the complex \lgA\ which is compatible with $D$. Therefore we 
obtain an equivariant 
$\IZ/2$ graded complex. The space of morphisms in the orbifold category 
$\CC_{W,\rho}$ is given by 
the $G$-fixed part of the cohomology groups $H^i(\op,\oq)$. 
In this way we obtain a triangulated category $\CC_{W,\rho}$. The shift functor
and the distinguished 
triangles can be constructed by imposing equivariance conditions on equations 
\shift,\
\triangleA,\ and \triangleB.\ 

\newsec{Fractional Branes} 

From now on we restrict ourselves 
to quasihomogeneous  Landau-Ginzburg potentials $W(X)$ of the form 
\eqn\lgD{ 
W(X) = X_0^{d_0} + X_1^{d_1} + \ldots + X_n^{d_n}}
where all $d_a\geq 3$, $a=0,\ldots,n$. 
The discrete symmetry group is $G=\IZ_d$, 
where $d=\hbox{l.c.m}(d_0,\ldots,d_n)$, and the action $\rho$ 
is specified by 
\eqn\lgE{ 
\rho(\omega)(X_0,\ldots,X_N) = (\omega^{w_0}X_0, \ldots, \omega^{w_n} X_n)} 
with $w_a={d\over d_a}$, $a=0,\ldots,n$.
This theory is equivalent to a $\IZ/d$ orbifold of a product of $n+1$ $(2,2)$ minimal models at levels 
$k_a = d_a-2$, in which D-branes can be explicitly 
described as rational boundary states satisfying Cardy's consistency condition \refs{\RS,\GS}. 
In particular ${\bf B}$-type boundary states are classified by a vector 
$L=(L_0, \ldots, L_n)$ with integer entries $0\leq L \leq d$ and an extra quantum number $M$ 
which takes even integer values $M\in \{0,2,4,\ldots,2d-2\}$. The orbifold theory has a quantum 
$\IZ/d$ symmetry which leaves $L$ invariant and shifts $M$ by two units $M\ra M+2$. 
The $L=0$ boundary states are known as fractional branes and play a special role in the 
context of Calabi-Yau compactifications, as discussed in the next section.

The goal of the present section is to find a relation between the algebraic approach explained above 
and the boundary state construction. In particular, we would like to know if there is a natural 
algebraic construction of the fractional boundary states described in the last paragraph.
In order to answer this question, let us start with the algebraic realization of ${\bf B}$-type 
boundary states in the one variable case.

\subsec{One variable models} 

Consider a LG potential $W=X^d$, $d\geq 3$. 
In the absence of the orbifold projection, the D-brane category $\CC_W$ 
has a very simple description \refs{\O,\KLIII}. 
The objects are pairs $\om_l$ of rank one $\IC[X]$-modules 
labeled by an integer $l\in \{1,\ldots,d-1\}$ with $m_1=X^l$, $m_0=X^{d-l}$. 
Factorizations that correspond to $m_1=1$ or $m_1=X^d$ are trivial objects in the category.
By construction, $\om_{d-l}$ is isomorphic to $\om_l[1]$, therefore $\om_l$ and $\om_{d-l}$ form a brane/anti-brane pair. Therefore we can restrict our attention to the range $l\le \left[{d\over 2}\right]$. 
\ifig\EndOfM{Endomorphisms of the object $\om_l$}{\epsfxsize2.0in\epsfbox{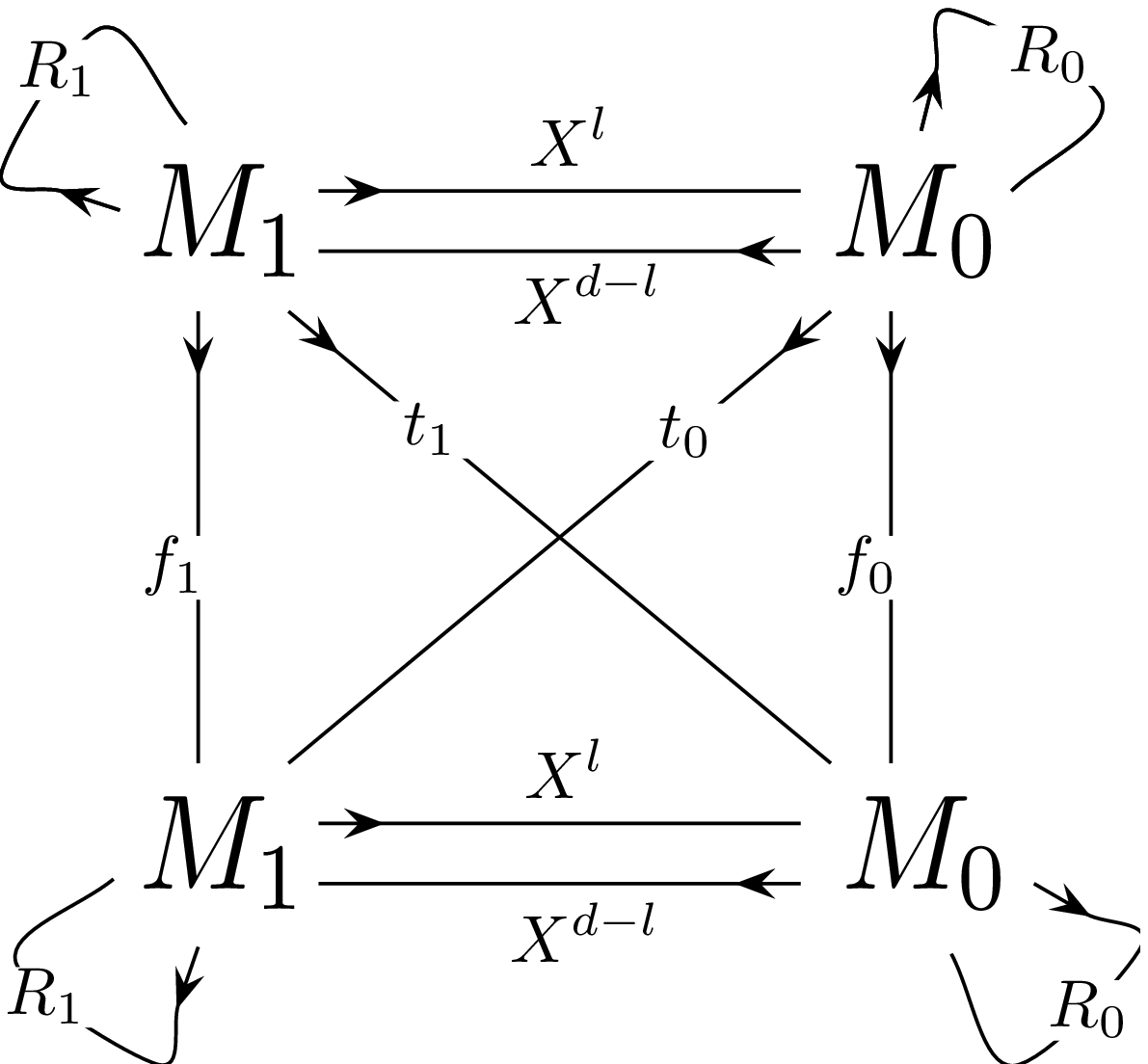}}
Let us study the morphisms between any pair of objects in $\CC_W$. Picking the standard two dimensional representation of \anticom, the generic morphism $\Phi$ in \morphism\ is written as
$$
\Phi=\left(\matrix{f_0 & t_1 \cr t_0 & f_1}\right)\ ,
$$
where we follow the conventions of \BHLS. A straightforward analysis of the $\IZ/2$ graded complex \lgA\ shows that 
\eqn\onevarA{ 
H^0(\om_l,\om_k) =\left\{\matrix{\IC[X]/(X^{\hbox{ min}\{k,d-l\}}), &\hbox{if}\ k\le l\cr
 \cr
\IC[X]/(X^{\hbox{min}\{l,d-k\}}), &\hbox{if}\ l\le k.\cr}\right.}
Moreover in the first case we can choose cohomology representatives of the form 
$f_0 = X^a,\ f_1 = X^{l-k}f_0$, $a=1,\ldots, \hbox{min}\{k,d-l\}$, while in the second case 
we can choose representatives $f_1 = X^a,\ f_0(X) =X^{k-l}f_1$, 
$a=1,\ldots,\hbox{min}\{l,d-k\}$. \
For fermionic morphisms we find similarly 
\eqn\onevarB{
H^1(\om_l,\om_k) =\left\{\matrix{\IC[X]/(X^{\hbox{min}\{k,l\}}),\hfill
 &\hbox{if}\ k\le d-l\cr  \cr
\IC[X]/(X^{\hbox{min}\{d-k,d-l\}}),&\hbox{if}\ k\ge d-l.\cr}\right.}
In the first case, we can choose representatives $t_1=X^a,\ t_0=-X^{d-l-k}t_1$, 
$a=1,\ldots,\hbox{min}\{k,l\}$ and in the second case $t_0=X^a,\ t_1=-X^{l+k-d}$, 
$a=1,\ldots,\hbox{min}\{d-k,d-l\}$.

Now we move to the orbifold category $\CC_{W,\rho}$. As discussed above, the objects 
are given by equivariant triples $(\op,R_0,R_1)$.
The $\IZ/d$ action on $\IC$ is given by $\rho(\omega)(X) = \omega X$, where 
$\omega=e^{{2\pi i \over d}}$. It suffices to consider the rank one objects $\op_l$
which generate the unorbifolded category. 
Then the representations $R_1,R_0$ are specified by two integers $\alpha_0, \alpha_1\in \{0,\ldots, d-1\}$. 
The equivariance condition \lgC\ yields $\alpha_0=\alpha_1+l$. 
Thus the rank one objects of $\CC_{W,\rho}$ are given by a pair of integers 
$(l,\alpha)$ so that $l\in\{1\ldots d-1\}$ and $\alpha\in\{0\ldots d-1\}$. 
We will denote such an object by $\om_{l,\alpha}$. Since the rank one objects generate $\CC_W$, 
the same will be true for $\CC_{W,\rho}$.
 
Let us determine the bosonic and fermionic morphisms between any two objects 
$\om_{l,\alpha}, \om_{k,\beta}$. We have to take the $G$-fixed part of the cohomology of \lgA.\ 
In the present case, this implies that the space of bosonic morphisms is generated by polynomial 
maps $f_0, f_1$ satisfying the equivariance conditions 
\eqn\bmorph{\eqalign{
f_0(X) \omega^{\alpha+l} &= \omega^{\beta+k} f_0(\omega^{-1}X)\cr
f_1(X) \omega^\alpha &= \omega^\beta f_1(\omega^{-1}X).}} 
\noindent
For maps of the form $f_0=X^a$ and $f_1=X^b$, the equations \bmorph,\ lead to the 
following conditions
\eqn\bmorphB{\eqalign{
a = k-l+b &= (\beta-\alpha)+(k-l) \qquad \hbox{for} \qquad 0\le a \le \hbox{min}\{k-1,d-l-1\}.}}
We can repeat the above analysis for fermionic morphisms, taking $t_0=X^{a'}$ and $t_1=-X^{b'}$. 
In this case, we find the conditions
\eqn\fmorph{
a' = d-(l+k)+b' = l+(\beta-\alpha) \qquad \hbox{for} \qquad 0\le a'\le \hbox{min}\{d-l-1,d-k-1\}.}
Therefore we find the following spaces of bosonic and respectively fermionic morphisms 
\eqn\znbmorph{
H^0(\om_{l,\alpha},\om_{k,\beta})= \left\{\matrix{ \IC,\ & \hbox{if}\ k\le l\ \hbox{and}\
 l-k\le \beta-\alpha \le \hbox{min}\{l-1, d-k-1\} \hfill \cr \cr
\IC,\ & \hbox{if}\ 
k\ge l\ \hbox{and}\ 0\le \beta-\alpha \le \hbox{min}\{l-1,d-k-1\}\hfill \cr \cr
0,\ &\hbox{otherwise}.\hfill \cr}\right.}
\eqn\znfmorph{
H^1(\om_{l,\alpha},\om_{k,\beta})= \left\{\matrix{
\IC,\ & \hbox{if}\ 
k \ge d-l\ \hbox{and}\ l \le \beta-\alpha \le \hbox{min}\{d+l-k-1,d-1\}\hfill \cr \cr
\IC,\ &\hbox{if}\
k \le d-l\ \hbox{and}\quad d-k \le \beta-\alpha \le \hbox{min}(d+l-k-1,d-1)\hfill \cr \cr
0,\ & \hbox{otherwise}. \hfill \cr}\right.}
where $l \le \left[{d\over 2}\right]$.
Note that the transformation $(l,\alpha)\rightarrow (d-l,\alpha+l)$ 
exchanges the bosonic and fermionic spectrum. 
Therefore $\om_{d-l,\alpha+l}$ is again isomorphic to the antibrane 
$\om_{l,\alpha}[1]$ of $\om_{l,\alpha}$.
For future reference, let us also note that the intersection matrix of $l=1$ states 
$\chi(\om_{1,\alpha},\om_{1,\alpha})$ is $(1-G^{-1})$, where $G$ is the shift matrix 
defined by the linear transformation $G:\om_{1,\alpha}\ra \om_{1,\alpha+1}$. 
\ifig\quiver{Quiver for the $\IZ_d$ orbit of $l=1$ states}{\epsfxsize2.5in
\epsfbox{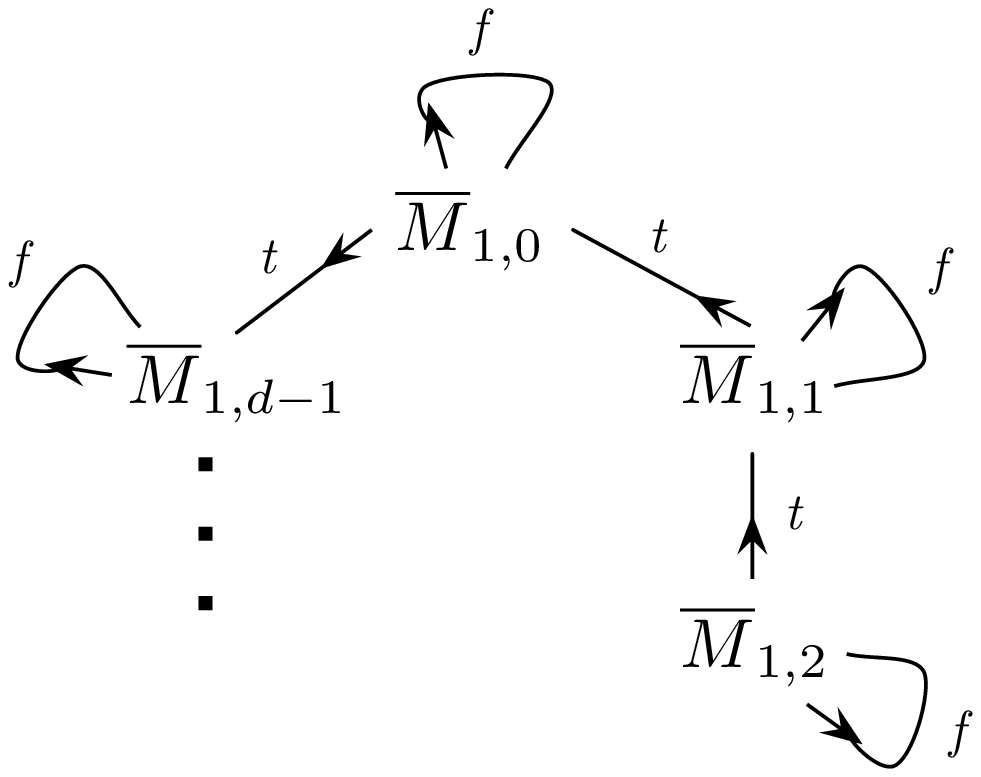}}
\noindent This is summarized in the quiver diagram \quiver\ .
One can extend the results found here to more general models of the form $W=X^{wd}$, 
where $w$ is an arbitrary integer, subject to the orbifold action 
$X\ra e^{2i\pi\over d} X$. Although we will not give the full details, note that the intersection 
matrix of $l=1$ objects becomes $(1-G^{-w})$ in this case.

\subsec{Comparison with Minimal Models}

Let us compare the above results to the rational boundary state construction in the
$\IZ/d$ orbifold of the $A_{d-2}$ minimal model. 
The boundary states are labeled by three quantum numbers $L\in \{0,\ldots, d-2\}$, 
$M \in \{-(d-1),\ldots, d\}\ \hbox{mod}\ 2d$, and 
$S \in \{-1,0,1,2\}\ \hbox{mod}\ 4$
subject to the constraint 
\eqn\bdsstA{
L+M+S = 0 \quad \hbox{mod}\quad 2.}
and the field identification
\eqn\bdstB{
(L,M,S) \sim (d-2-L,M+d,S+2).}
Topologically twisted ${\bf B}$-branes correspond to Ramond sector boundary states, 
which are characterized by $S\in\{-1,1\}$.
Moreover, the transformation $(L,M,S) \ra (L,M,S+2)$ maps a brane to its antibrane. 
Therefore, using the equivalence relations \bdstB,\ we can label rational boundary states 
by $\left.|L,M\right\rangle$, adopting the convention that $\left.|L,M\right\rangle$
and $\left.|d-2-L,M+d\right\rangle$ form a brane/anti-brane pair for any $(L,M)$. 

We have an intersection pairing on the set of boundary states defined by the open string Witten 
index, which counts open string Ramond ground states with sign \refs{\DF}.  
This pairing can be evaluated using CFT techniques, obtaining \refs{\HIV}
\eqn\mmintA{\eqalign{
& I(\left.|L_1,M_1,S_1\right\rangle,\,\left.|L_2,M_2,S_2\right\rangle)= 
(-1)^{{(S_2-S_1)\over 2}} N_{L_1,L_2}^{M_2-M_1} \cr
& N_{L_1,L_2}^{M_2-M_1} = \left\{\matrix{1,& \hbox{if}\ |L_1-L_2|\le M_2-M_1\le 
\hbox{min}\{L_1+L_2,\,2d-4-L_1-L_2\}\cr  0, &\hbox{otherwise .}\hfill \cr}\right.\cr}}
Using the equivalence relation \bdstB\ as explained in the previous paragraph, we can rewrite 
this formula as follows 
\eqn\mmintB{
I(\left.|L_1,M_1\right\rangle,\,\left.|L_2,M_2\right\rangle)= 
\left\{\matrix{1,\hfill &\hbox{if}\ |L_1-L_2|\le M_2-M_1 \le \hfill \cr & \qquad\qquad 
\hbox{min}\{L_1+L_2,\,2d-4-L_1-L_2\}\hfill\cr \cr 
-1, & \hbox{if}\ |L_1+L_2-d+2|\le M_2-M_1-d\le \hfill \cr
& \qquad \qquad \hbox{min}\{L_1-L_2+d-2,L_2-L_1+d-2\}\hfill \cr \cr
0,\hfill & \hbox{otherwise .}\hfill \cr}\right.}

In order to find a map between these boundary states and the objects constructed earlier, 
recall that 
the $\IZ/d$ orbifold has a $\IZ/d$ quantum symmetry which acts on boundary states by shifting 
$M\ra M+2$, leaving $(L,S)$ fixed. In the D-brane category, the same quantum symmetry maps a 
brane $\om_{l,\alpha}$ to $\om_{l,\alpha+1}$. Given the range of $L\in \{0,\ldots, d-2\}$
and respectively $l\in \{1,\ldots, d-1\}$, we are lead to the following identifications
\eqn\mapA{
L = l-1 \qquad \hbox{and}\qquad M = 2\alpha+l.}
As a first check, note that this identification maps brane/anti-brane pairs to brane/anti-brane 
pairs. Furthermore, one can check by simple computations that the intersection pairing 
\mmintB\ agrees with the previous results \znbmorph,\ \znfmorph\ under the map \mapA.\
In particular the intersection matrix of the $L=0$ boundary states agrees with the 
result found before for $\om_{1,\alpha}$, namely $(1-G^{-1})$. This correspondence can be easily extended 
to $\IZ/d$ orbifolds of $A_{dw-2}$ minimal models, where $w\geq 1$. In this case, the intersection 
matrix is $(1-G^{-w})$, in agreement with LG results. 
In order to extend this correspondence to more general LG orbifolds, 
we need an algebraic construction which will be described next. 

\subsec{Tensor Product}  

In more general situations we have to construct matrix factorizations for LG superpotentials 
depending on several variables $X_a$, $a=1,\ldots,n$. This is a more difficult task
than the one variable case. In the following we will describe a systematic approach 
to this problem based on a tensor product construction. This is the algebraic counterpart 
of tensoring boundary states in products of minimal models. 

Let us start with a simple example, namely a superpotential $W(X_1,X_2)= W_1(X_1) + W_2(X_2)$
depending on two variables. We first present the construction in terms of boundary 
couplings using the formalism of section 2. 
Consider two separate {\bf B}-type boundary conditions for $W_1, W_2$ 
specified by boundary couplings of the form $\Pi_1F_1(X_1)$ and respectively 
$\Pi_2F_2(X_2)$ satisfying 
\eqn\bcondA{
F_1(X_1)G_1(X_1)=W_1(X_1),\qquad F_2(X_2)G_2(X_2)=W_2(X_2).}
These boundary conditions correspond to rank one factorizations $\op, \oq$ of 
$W_1,W_2$. 
The sum $\int dx^0\,d\theta\,\left[\Pi_1F_1(X_1)+ \Pi_2F_2(X_2)\right]$
determines a supersymmetric boundary condition for $W$, as explained in section 2. 
In order to translate this construction in algebraic language, pick the standard four dimensional representation of the Clifford algebra generated by $\pi_1,\pi_2,{\bar \pi}_1, {\bar \pi}_2$. 
Then the boundary supercharge takes the form 
\eqn\matfact{
D = 
\left[\matrix{0 & 0 & F_1(X_1) & F_2(X_2) \cr 
0 & 0 & G_2(X_2) & -G_1(X_1) \cr 
G_1(X_1) & F_2(X_2) & 0 & 0\cr 
G_2(X_2) & -F_1(X_1) & 0 & 0 }\right].}
The corresponding object in the category $\CC_W$ is of the form 
\eqn\obj{\eqalign{
& \qquad \qquad \qquad \quad \xymatrix{
\IC[X_1,X_2]^{\oplus 2}\ar@<0.6ex>[r]^{r_1} 
&\IC[X_1,X_2]^{\oplus 2} \ar@<0.6ex>[l]^{r_0}}\cr 
& r_1= \left[\matrix{F_1(X_1) & F_2(X_2) \cr G_2(X_2) & -G_1(X_1) \cr }\right],\quad 
r_0=\left[\matrix{G_1(X_1) & F_2(X_2) \cr G_2(X_2) & -F_1(X_1) \cr}\right].\cr}}
We will denote it by $\op\otimes \oq$. Consider two such tensored  objects 
$\op_1\otimes \oq_1$ and $\op_2\otimes \oq_2$, where $\op_i$ are objects in 
$\CC_{W_1}$ and $\oq_i$ in $\CC_{W_2}\,$. The spaces of morphisms between them 
can be found by analyzing the boundary chiral ring in the associated open string sector. 
From the form of \bdrycharge\ we see that any element in the cohomology of $\CD$ 
must have the form 
\eqn\productmorph{
\Phi=\prod_{a=1,2}\Phi_{a}
}
where $\Phi_{a}$ is a physical operator in the LG theory with superpotential $W_{a}(X_{a})$. It follows that
\eqn\homplushom{
H^k(\op_1\otimes \oq_1,\op_2\otimes \oq_2) = 
\mathop{\bigoplus_{i,j=0,1}}_{i-j\equiv k(2)} H^i(\op_1,\op_2)\otimes H^j(\oq_1, \oq_2).} 
This is in fact a special case of a more general construction, which is best 
described in the algebraic framework. 

Let us now consider a superpotential $W=W(X_a, Y_b)$, $a=0,\ldots, n$, 
$b=0,\ldots, m$ 
which can be written as a sum 
\eqn\sumA{
W(X_a,Y_b) = W_1(X_a) + W_2(Y_b).}
For simplicity we will use the notation $X=(X_a)_{a=1,\ldots,n}$, 
$Y=(Y_b)_{b=1,\ldots,m}$. 
Let $\op$, $\oq$ be two arbitrary matrix factorizations of $W_1(X)$ and 
respectively 
$W_2(Y)$. We claim that one can form a canonical matrix factorization $\op\otimes \oq$ of $W(X_a,Y_b)$ 
as follows. 
Recall that $P_0,P_1$ are free $\IC[X]$-modules of arbitrary rank, and, 
similarly, 
$\oq_0,\oq_1$ are free $\IC[Y]$-models. 
Note that we have standard ring morphisms $\IC[X],\IC[Y] \ra\, \IC[X,Y]$ 
corresponding to the projections $\pi_1:\IC^{n+m} = \IC^n\times \IC^m \ra\,
 \IC^n$ 
and respectively $\pi_2: \IC^{n+m} = \IC^n\times \IC^m \ra\, \IC^m$. 
For any $\IC[X]$-module $A$, we have a pull-back $\IC[X,Y]$-module 
$\pi_1^*A = A\otimes_{\IC[X]}\IC[X,Y]$. 
Similarly, any $\IC[Y]$-module $B$ gives rise to a $\IC[X,Y]$-module 
$\pi_2^* B = B \otimes_{\IC[Y]}\IC[X,Y]$. 
We take 
\eqn\tensorA{\eqalign{ 
& (\op\otimes \oq)_1 = \pi_1^* P_1 \otimes_{\IC[X,Y]} \pi_2^* Q_0 \oplus 
\pi_1^*P_0 \otimes_{\IC[X,Y]} \pi_2^* Q_1\cr
& (\op\otimes \oq)_0 = \pi_1^*P_0\otimes_{\IC[X,Y]} \pi_2^* Q_0 \oplus \pi_1^* P_1 \otimes_{\IC[X,Y]} \pi_2^* Q_1.\cr}}
The maps 
$\xymatrix{(\op\otimes \oq)_1 \ar@<1ex>[r]^{r_1}&(\op\otimes \oq)_0 \ar@<1ex>[l]^{r_0}\\}$
are given by 
\eqn\tensorB{ 
r_1=\left[\matrix{ p_1\otimes 1 & 1\otimes q_1 \cr 1\otimes q_0 & -p_0\otimes 1\cr}\right]
\qquad 
r_0=\left[\matrix{ p_0\otimes 1 & 1\otimes q_1 \cr 1\otimes q_0 & -p_1\otimes 1\cr}\right].}
It is straightforward to check that this is a matrix factorization of $W(X,Y)=W_1(X)+W_2(Y)$. 

Next, we would like to determine the spaces of morphisms between two tensor product objects 
$\op_1\otimes \oq_1$, $\op_2\otimes \oq_2$. The most efficient way to proceed is by 
reformulating the above construction in terms of differential 
$\IZ/2$ graded modules, as explained 
below \windex.\ The objects $\op$, $\oq$ considered in the previous paragraph can be regarded
as differential $\IZ/2$ graded modules 
 $(P,\eta_P,p)$, $(Q,\eta_Q, q)$ satisfying the conditions 
\eqn\triplesA{\eqalign{ 
& \eta_P\, p +p\,\eta_P =0,\qquad p^2=W_1(X),\qquad \eta_P^2 = 1\cr
& \eta_Q\, q +q\,\eta_Q =0,\qquad q^2=W_2(Y),\qquad \eta_Q^2 = 1.\cr}}
The tensor product $\op\otimes \oq$ corresponds to the triple 
\eqn\tensprod{
\left(\pi_1^*P\otimes \pi_2^*Q, \eta_P\otimes \eta_Q, p\otimes \eta_Q + 1\otimes q\right).}
Now, the spaces of morphisms $H^{0,1}(\op_1,\op_2)$ and respectively $H^{0,1}(\oq_1, \oq_2)$ are determined by the differential $\IZ/2$ graded modules 
$\left(\hbox{Hom}(\op_1, \op_2), \eta_{P_1,P_2}, D_1\right)$ and respectively 
$\left(\hbox{Hom}(\oq_1, \oq_2), \eta_{Q_1,Q_2}, D_2\right)$
where 
\eqn\triplesB{\eqalign{ 
& \eta_{P_1, P_2} D_1 + D_1 \eta_{P_1, P_2} =0, \qquad D_1^2=0,\qquad  \eta_{P_1,P_2}^2 = 1\cr
& \eta_{Q_1, Q_2} D_2 + D_2 \eta_{Q_1, Q_2} =0, \qquad D_2^2=0, \qquad \eta_{Q_1,Q_2}^2 = 1.\cr
}}
The differential $\IZ/2$ graded $\IC[X,Y]$-module which determines the morphism  spaces between $\op_1\otimes \oq_1$, $\op_2\otimes \oq_2$ is then given by 
\eqn\triplesC{ 
\left(
\hbox{Hom}(\op_1, \op_2)\otimes_{\IC[x,y]} \hbox{Hom}(\oq_1, \oq_2), \eta_{P_1,P_2}\otimes 
\eta_{Q_1,Q_2}, 
D_1\otimes \eta_{Q_1,Q_2}+1\otimes D_2\right).}  
It is straightforward to check that the data \triplesC\ forms a differential $\IZ/2$-graded 
module, using \triplesB.\ 
This is in fact a familiar construction in homological algebra, namely the tensor product of two 
$\IZ/2$ graded differential complexes. Then we can use the algebraic K\"unneth formula to relate  the 
cohomology of $\triplesC$ to that of the individual complexes 
 $(P,\eta_P,p)$, $(Q,\eta_Q, q)$. This yields an exact sequence of $\IC[X,Y]$-modules 
\eqn\tensorcoh{ \eqalign{
0 & \ra \mathop{\bigoplus_{i,j=0,1}}_{i-j\equiv k (2)} H^i(\op_1,\op_2) \otimes H^j(\oq_1,\oq_2) 
\,\ra \,H^{k}( \op_1\otimes \oq_1, \op_2\otimes \oq_2)\cr & \ra 
\mathop{\bigoplus_{i,j=0,1}}_{i-j\equiv k-1(2)} \hbox{Tor}_1^{\IC[X,Y]}(H^i(\op_1,\op_2), 
H^j(\oq_1,\oq_2))\,\ra \,0.}}
In order to compute the $\hbox{Tor}_1$ 
group in the third term of \tensorcoh,\ we have to pick a locally free resolution 
$\CF^{\cdot}\ra H^j(\oq_1,\oq_2)$ and construct the complex 
\eqn\torA{ 
0\ra \pi_2^*\CF^{\cdot} \otimes_{\IC[x,y]} \pi_1^*H^i(\op_1,\op_2)\,.}
The group $\hbox{Tor}^1_{\IC[X,Y]}(H^i(\op_1,\op_2), 
H^j(\oq_1,\oq_2))$ is the first cohomology group of this complex. 
We claim that this is always zero because the complex \torA\ is exact. To justify 
this claim, note that $\pi_2^*\CF$ is always exact since $\IC[X,Y]$ is a flat $\IC[Y]$-module. 
Moreover, the differentials of $\pi_2^*\CF$ are pulled back from $\CF$. Then one can check 
by direct computations that such a complex will remain exact after tensoring by the pull-back 
module $\pi_1^*H^i(\op_1,\op_2)$.  
Therefore we obtain the following simple formula 
\eqn\tensormorph{ 
 H^{k}( \op_1\otimes \oq_1, \op_2\otimes \oq_2) =\mathop{\bigoplus_{i,j=0,1}}_{i-j\equiv k (2)} 
H^i(\op_1,\op_2) \otimes H^j(\oq_1,\oq_2).}

The tensor product can be easily extended to orbifold categories. 
Consider a finite cyclic group $G$ and representations 
$\rho_1:G\ra GL(n+1, \IC)$, $\rho_2:G\ra GL(m+1, \IC)$. There is an 
obvious induced representation $(\rho_1, \rho_2):G \ra GL(n+m+2,\IC)$. 
In the orbifold categories $\CC_{W_1,\rho_1}$, $\CC_{W_2,\rho_2}$
we have to specify representations of $G$ on the pairs $\op, \oq$ 
as described in section 3.2. The objects are triples 
$(\op,R_1,R_0)$, $(\oq,S_1,S_0)$ satisfying the equivariance condition 
\lgC.\ In order to produce objects of $\CC_{W_1+W_2, (\rho_1,\rho_2)}$, 
it suffices to specify a group action and impose equivariance conditions on tensor 
products of the form $\op\otimes \oq$. 

Alternatively, one can easily construct such objects by taking tensor products 
of equivariant triples $(\op,R_1,R_0)$, $(\oq,S_1,S_0)$. 
The representations $R_1,R_0,S_1,S_0$ induce canonical representations 
of $G$ on the tensor product modules \tensorA\ so that the morphisms \tensorB\ 
satisfy the equivariance condition \lgC.\ Therefore we obtain an object 
$(\op,R_1,R_0)\otimes (\oq,S_1,S_0)$ of $\CC_{W,(\rho_1,\rho_2)}$. 
The morphisms between two such objects can be determined by imposing 
$G$-invariance in the formula \tensormorph.\ 
Note that for fixed $\op, \oq$, the tensor products 
$(\op,R_1,R_0)\otimes (\oq,S_1,S_0)$
are not in $1-1$ correspondence with the quadruples $(R_0, R_1, S_1, S_0)$. 
Two different quadruples $(R_0, R_1, S_1, S_0)$ may result in isomorphic
tensor products. In order to clarify the details, we return to the construction of fractional branes. 

\subsec{Fractional Branes} 

We consider an orbifolded LG model with a quasihomogeneous superpotential 
\eqn\fractA{W(X) = X_0^{d_0} + X_1^{d_1} + \ldots + X_n^{d_n}}
with $d_a\geq 3$, $a=0,\ldots,n$ that corresponds to a  
$\IZ/d$ orbifold of a product of minimal models at levels 
$k_a = d_a-2$. We will focus on the $\IZ/d$ orbit of rational boundary states with $L=0$. These are essentially constructed by tensoring rational boundary states with $L_a=0$ and arbitrary values of $M_a$. One can show that two boundary states with the same $L$ and total quantum number $M=\sum_{a=0}^n {M_a}$ are isomorphic. We will show here that the algebraic counterpart of this construction is the tensor product introduced in the previous subsection. 

Let us start again with a two variable example of the form 
\eqn\fractB{W(X) = X_1^{d_1} + X_2^{d_2}.} 
Let $W_1(X_1)=X_1^{d_1}$, $W_2(X_2) = X_2^{d_2}$. The $\IZ/d$-orbifold action is given by 
$X_1 \ra \omega^{w_1} X_1$, $X_2\ra \omega^{w_2} X_2$, where $d=\hbox{l.c.m}\{d_a\}$ 
and $w_a=d/d_a$, $a=1,2$. 
To each pair $\om_{l_1}$, $\om_{l_2}$ of rank one factorizations of $W_1(X_1)$,$W_2(X_2)$ 
we can associate the tensor product $\om_{l_1l_2} = \om_{l_1} \otimes \om_{l_2}$, which
is a rank two matrix factorization of $W(X)$. 
To define fractional branes in the orbifold theory, 
we have to specify a two dimensional representation of the orbifold group $\IZ/d$ 
on $\om_{l_1l_2}$. 
After imposing the equivariance conditions \lgC\ we find a collection of $d$ objects 
labeled by an integer $\mu\in\{0,\ldots,d-1\}$ which are cyclicly permuted by the quantum 
$\IZ/d$ symmetry of the orbifold theory. The corresponding representations on 
$(\om_{l_1}\otimes\om_{l_2})_0$ and 
$(\om_{l_1}\otimes\om_{l_2})_1$ are  
\eqn\tensrep{
R_0(\omega)=\left(\matrix{\omega^{\mu} & 0 \cr 0 & \omega^{\mu+l_1+l_2}}\right) \qquad
R_1(\omega)=\left(\matrix{\omega^{\mu+l_1} & 0 \cr 0 & \omega^{\mu+l_2}}\right).}
The same result can be obtained by directly tensoring the 
objects $\om_{l_1,\alpha_1}$ $\om_{l_2,\alpha_2}$ of the one variable orbifold theories  
provided that $\mu=\alpha_1+\alpha_2$. 
As noted in the last paragraph of the previous subsection, 
the tensor product depends only on the sum $\mu=\alpha_1+\alpha_2$, 
not on the individual values $\alpha_1, \alpha_2$. 

In order to find the morphisms between $\om_{l_1l_2,\mu}$, $\om_{k_1k_2,\mu'}$ 
we have to take the $G$-invariant part of 
\eqn\fractC{ 
H^k(\om_{l_1l_2},\om_{k_1k_2})= \mathop{\bigoplus_{i,j=0,1}}_{i-j\equiv k(2)}H^i(\om_{l_1},\om_{k_1})
\otimes H^j(\om_{l_2}, \om_{k_2})}
under the $G$-action determined by $(\rho_1,\rho_2),\mu,\mu'$. More precisely, we have to identify the 
trivial $G$-module in the direct sum decomposition of $H^k(\om_{l_1l_2},\om_{k_1k_2})$ into 
irreducible $G$-modules. Equivalently, we can consider the $G$-action on $H^k(\om_{l_1l_2},\om_{k_1k_2})$
induced by $\rho$ and the trivial representations on $\om_{l_1l_2}, \om_{k_1k_2}$, and identify the 
$\mu'-\mu$ block in its decomposition into irreducible $G$-modules. 
Taking this point of view,  
consider the $G$-action on the right hand side of \fractC\ 
induced by $\rho_1, \rho_2$ and the trivial representations on $\om_{l_1},\ldots, \om_{k_2}$. 
An important observation is that the irreducible $G$-modules in the decomposition 
of $H^i(\om_{l_1},\om_{k_1})$ are isomorphic to the morphism spaces $H^i(\om_{l_1,0}, \om_{k_1,\alpha})$
in the orbifold category $\CC_{W_1,\rho_1}$. Therefore we have  
\eqn\fractD{\eqalign{ 
& H^i(\om_{l_1},\om_{k_1}) = \bigoplus_{\alpha=0}^d H^i(\om_{l_1,0}, \om_{k_1,\alpha}) \cr
& H^j(\om_{l_2},\om_{k_2}) = \bigoplus_{\beta=0}^d H^j(\op_{l_2,0}, \op_{k_2,\beta}).\cr}}
The $\mu'-\mu$ block in the decomposition of the right hand side of \fractC\ is therefore 
\eqn\fractE{ \eqalign{
& H^k(\om_{l_1l_2,\mu},\om_{k_1k_2,\mu'}) = \cr
&\qquad 
\mathop{\bigoplus_{i,j=0,1}}_{i-j\equiv 0(2)}
\bigoplus_{\alpha,\beta=0}^d
\left[H^i(\om_{l_1,0},\om_{k_1,\alpha})\otimes 
H^j(\om_{l_2,0},\om_{k_2,\beta})\right]
\cdot\delta_{\mu'-\mu-(\alpha+\beta)}.\cr}}
This is a very useful formula expressing 
the morphism spaces in $\CC_{W_1+W_2,(\rho_1,\rho_2)}$ in terms of 
morphism spaces in $\CC_{W_1,\rho_1}$, $\CC_{W_2,\rho_2}$. 
A direct consequence of \fractE\ is a similar relation between intersection numbers
\eqn\fractF{ 
\chi(\om_{l_1l_2,\mu},\om_{k_1k_2,\mu'}) = 
\sum_{\alpha,\beta=0}^d \chi(\om_{l_1,0}, \om_{k_1,\alpha})\chi(\om_{l_2,0}, 
\om_{k_2,\beta})\cdot\delta_{\mu'-\mu-(\alpha+\beta)}.}

Our goal is to show that there is a one to one correspondence between the $L=0$ boundary states 
in the minimal model theory and the tensor product objects $\om_{1,1,\mu}$. Obviously, they 
have the same transformation properties under the orbifold quantum symmetry. Moreover, one can check 
as in the one variable case that such an identification is consistent with mapping 
branes to anti-branes. The main test of this proposal is the comparison of intersection matrices. 
The expected CFT answer for rational boundary states is $\prod_{a=0}^n(\II-G^{w_a})$. 
Equation \fractF\ gives 
\eqn\fractG{
\chi(\om_{1,1,\mu},\om_{1,1,\mu'}) = 
\sum_{\alpha,\beta=0}^d \chi(\om_{1,0}, \om_{1,\alpha})\chi(\om_{1,0}, 
\om_{1,\beta})\cdot \delta_{\mu'-\mu-(\alpha+\beta)}.}
Note that the intersection numbers 
$\chi(\om_{1,\alpha_1}, \om_{1,\alpha_2})$ are invariant under a simultaneous 
shift $\alpha_1\ra \alpha_1+r$, $\alpha_2\ra \alpha_2+r$. Then, using \fractE,\ 
we find that the intersection matrix 
of $\{\om_{1,1,\mu}\}$ is simply the product of the one variable intersection matrices 
\eqn\fractH{
\chi(\om_{1,1,\mu},\om_{1,1,\mu'}) = [(\II-G^{-w_1})(\II-G^{-w_2})]_{\mu\mu'}}
which is in exact agreement with the CFT result. This is very strong evidence for our proposal. 

Since the above discussion is somewhat abstract, let us construct the endomorphisms of $\om_{11}$
explicitly. Using \onevarA\ and \onevarB,\ we know that for the $W=X^d$ theory, 
the object $\om_1$ has one bosonic and one fermionic endomorphism. It follows from
\homplushom\ that $\om_{11}$ has two bosonic and two fermionic endomorphisms, 
independent of the value of $d$. The fermionic ones have the form 
\eqn\Tmatrices{\eqalign{
T_0 
&=(t_0^{(0)}\,\pib_0+t_1^{(0)}\,\pi_0)\,
(f_0^{(1)}\,\pi_1\pib_1+f_1^{(1)}\,\pib_1\pi_1)=\left(\matrix{
0 & 0 & 1 & 0 \cr
0 & 0 & 0 & X_0^{d-2} \cr
-X_0^{d-2} & 0 & 0 & 0 \cr
0 & -1 & 0 & 0 }\right)\cr
T_1 &= (f_0^{(0)}\,\pi_0\pib_0+f_1^{(0)}\,\pib_0\pi_0)\,
(t_0^{(1)}\,\pib_1+t_1^{(1)}\pi_1)=
\left(\matrix{
0 & 0 & 0 & 1 \cr
0 & 0 & -X_1^{d-2} & 0 \cr
0 & 1 & 0 & 0 \cr
-X_1^{d-2} & 0 & 0 & 0 }\right)
}}
where we have used the results of the one variable case. Similarly, the bosonic 
morphisms have the form
$$\eqalign{
\II_4 &= (f_0^{(0)}\,\pi_0\pib_0+f_1^{(0)}\,\pib_0\pi_0)\,
(f_0^{(1)}\,\pi_1\pib_1+f_1^{(1)}\,\pib_1\pi_1) \cr
T_0\cdot T_1 &= 
(t_0^{(0)}\,\pib_0+t_1^{(0)}\,\pi_0)\,(t_0^{(1)}\,\pib_1+t_1^{(1)}\,\pi_1) =
\left(\matrix{0 & 1 & 0 & 0 \cr 
-X_0^{d-2}X_1^{d-2} & 0 & 0 & 0 \cr 
0 & 0 & 0 & -X_0^{d-2}\cr 
0 & 0 & X_1^{d-2} & 0 }\right)
} $$
We know from our earlier discussion that in the orbifold theory, the 
intersection matrix for the fractional branes has the form 
\eqn\twom{
(\II-G^{-1})^2=\II-2\,G^{-1}+G^{-2},
}  
where $G$ is the shift matrix that corresponds to moving forward in the $\IZ_d$ orbit by one unit. 
We can see this from the explicit form of the endomorphisms of $\om_{1,1}$. 
After taking the orbifold 
these become morphisms between different objects in the orbit $\om_{1,1,\mu}\,$, 
according to their $\IZ_d$ charge. The charges are determined by the group 
action on the objects, 
as in \tensrep.  For tensor products of $l=1$ objects, it is possible 
to see that each fermionic constituent contributes charge -1 to the total charge of the 
morphism obtained by tensor product. Thus, $T_0$ and $T_1$ have charges $-1$, 
while $T_0\cdot T_1$ has charge $-2\,$, in agreement with \twom. 

We can generalize this construction to arbitrary numbers of variables. 
By taking successive tensor products
we find objects of the form $\om_{l_0,\ldots,l_n,\mu}$ where $\mu=0,\ldots,d-1$. The morphism spaces 
and intersection numbers of two such objects can be computed by induction. 
For unorbifolded objects, we have $\om_{l_0,\ldots,l_{n-1},l_n} = \om_{l_0,\ldots,l_{n-1}} 
\otimes \om_{l_n}$ and similarly $\om_{k_0,\ldots,k_{n-1},k_n} = \om_{k_0,\ldots,k_{n-1}} 
\otimes \om_{k_n}$. Repeating the steps between \fractC\ and \fractD\ we find the following 
recursion formula for morphisms 
\eqn\fractI{ \eqalign{
& H^k(\om_{l_0,\ldots, l_n,\mu},\om_{k_0,\ldots,k_n,\mu'}) = \cr
& \mathop{\bigoplus_{i,j=0,1}}_{i-j\equiv 0(2)}
\bigoplus_{\alpha,\beta=0}^d
\left[H^i(\om_{l_0,\ldots,l_{n-1},0},\om_{k_0,\ldots,k_{n-1},\alpha})\otimes 
H^j(\om_{l_n,0},\om_{k_n,\beta})\right]
\delta_{\mu'-\mu-(\alpha+\beta)}.\cr}}
This yields a similar recursion formula for intersection numbers 
\eqn\fractJ{ \eqalign{
& \chi(\om_{l_0,\ldots, l_n,\mu},\om_{k_0,\ldots,k_n,\mu'}) = \cr
& \qquad 
\sum_{\alpha,\beta=0}^d \chi(\om_{l_0,\ldots,l_{n-1},0}, \om_{k_0,\ldots,k_{n-1},\alpha})
\chi(\om_{l_n,0}, 
\om_{k_n,\beta})\cdot\delta_{\mu'-\mu-(\alpha+\beta)}.\cr}}
By specializing \fractJ\ to the fractional branes $\om_{1,\ldots,1,\mu}$, we find again 
that the intersection matrix can be written as a product 
\eqn\fractK{
\chi\left(\om_{1,\ldots,1,\mu},\om_{1,\ldots,1,\mu'}\right) = [\prod_{a=0}^n(\II-G^{-w_a})]_{\mu\mu'},}
which is the expected CFT result. 

\newsec{New Fractional Branes and Geometric Interpretation} 

So far we have reproduced the known boundary state results from an algebraic point of view. 
In this section we construct a new class of fractional branes in homogeneous Landau-Ginzburg 
models which do not have a rational boundary state counterpart. 
We also find their geometric interpretation in some Calabi-Yau examples 
and show that they are not restrictions of bundles (or sheaves) on the 
ambient toric variety. It is worth noting that a special case of this construction 
yields a single D0-brane on the Fermat quintic. Landau-Ginzburg boundary conditions 
corresponding to non-rational boundary states have been previously considered in \GJcycles.\
Although that construction is also based on factorization of the superpotential, it is not clear 
how it is related to the present approach. 
Let us start with the building blocks of our construction. 

\subsec{Rank One Factorizations for Two Variable Models} 

The basic idea is quite straightforward. The fractional branes were constructed 
by taking tensor 
products of one variable rank one factorizations. However, in certain 
cases one can use 
alternative building blocks consisting of rank one factorizations of two 
variable models. Consider a homogeneous superpotential of the form 
\eqn\homsup{
W(X_0, X_1)= X_0^{d} + X_1^{d}.} 
It is clear that one can construct rank one factorizations of the form  
\eqn\rkoneA{
\op_{\eta}=\left(\xymatrix{P_1 \ar@<1ex>[r]^{p_1}&P_0 
\ar@<1ex>[l]^{p_0}\\}\right),\qquad
p_1=X_0-\eta X_1,\qquad 
p_0= \prod_{\eta'\neq \eta}^{n-1}(X_0-\eta' X_1)\,,}
where $\{\eta\}$ is a complete set of $d$-th roots of $-1$. 
Given two such factorizations $\op_\eta,\op_{\eta'}$, 
one can easily determine the morphism spaces 
\eqn\rkoneB{
H^0(\op_\eta,\op_{\eta'}) = \left\{\matrix{\IC[Y]/(Y^{d-1}),\
&\hbox{if}\ \eta=\eta' \hfill\cr 
0,\ \hfill & \hbox{if} \ \eta\neq \eta'\hfill\cr}\right.\qquad
H^1(\op_\eta,\op_{\eta'}) = \left\{\matrix{\IC,\ & \hbox{if}\ \eta\neq \eta'\hfill
\cr 0,\ \hfill &
\hbox{if}\ \eta=\eta'.\hfill \cr}\right.}
In the orbifold category, we obtain rank one objects of the form 
$\op_{\eta,\alpha}$,  $\alpha=0,\ldots,d-1$, where $\alpha$ specifies the 
action of $\IZ/d$ on $(\op_\eta)_1$. The action of the orbifold group on 
$(\op_\eta)_0$ follows from equivariance constraints. The morphism spaces  
can be found by imposing an equivariance condition on the morphisms \rkoneB.\
The result is 
\eqn\rkoneC{\eqalign{
& H^0(\op_{\eta,\alpha},\op_{\eta',\beta}) = \left\{\matrix{\IC,\ \hfill
&\hbox{if}\ \eta=\eta' \ \hbox{and}\ \beta \neq \alpha -1 \cr 
0,\ \hfill & \hbox{otherwise}\hfill \cr}\right.\cr
& H^1(\op_{\eta,\alpha},\op_{\eta',\beta}) = \left\{\matrix{\IC,\ & \hbox{if}\ \eta\neq \eta'\ 
\hbox{and}\ \alpha=\beta+1\cr 
0,\ \hfill & \hbox{otherwise .}\hfill \cr}\right.\cr}}
We can write the intersection matrix of two  
objects in any given $\IZ_d$ orbit as
\eqn\rkoneCAA{
\chi(\op_{\eta,\alpha}\,,\op_{\eta,\beta})=\left[\II+G+G^2+\ldots+G^{d-2}\right]_{\alpha\beta}}

We have shown in the previous section that the superpotential \homsup\ admits tensor product rank two
matrix factorizations $\om_{l_1,l_2}$.  For later applications, we need to determine the 
morphism spaces between the fractional branes
$\om_{1,1}$ and the rank one objects $P_\eta$. This analysis is performed in appendix A 
where we find 
\eqn\rkoneCA{
H^0(\om_{1,1},\op_\eta)=\IC,\qquad H^1(\om_{1,1},\op_\eta)=\IC.}
The morphisms spaces between orbifold objects are given by 
\eqn\rkoneCB{ 
H^0(\om_{1,1,\,\mu'},\op_{\eta,\mu'}) = \left\{\matrix{ \IC,\ \hfill & 
\hbox{if}\ \mu'=\mu-2 \cr
0,\ \hfill & \hbox{otherwise} \hfill\cr}\right.,\qquad 
H^1(\op_{\eta,\mu}, \om_{1,1,\mu'}) = \left\{\matrix{ \IC, \ \hfill & 
\hbox{if} \ \mu'=\mu-1\cr 0,\ \hfill & \hbox{otherwise .}\hfill \cr}\right.}
The intersection matrix is thus
\eqn\rkoneCC{ 
\chi(\om_{1,1,\mu},\op_{\eta,\mu'}) = (G^{-2}-G^{-1})_{\mu\mu'}\,,}
where $G$ is the shift matrix introduced in section 3.2. This result holds for any value of $\eta$. 

\subsec{New Fractional Branes} 

Using the above rank one factorizations, we can construct new D-branes in LG orbifold models 
defined by homogeneous superpotentials  
\eqn\homsupB{ 
W(X)=X_0^d+\ldots+X_n^d}
as follows. Recall that the fractional branes were constructed by taking tensor products of rank 
one factorizations associated to the monomials $X_a^d$ in \homsupB.\ 
In order to obtain more general objects we can decompose $W(X)$ as a sum of monomials 
$W_a=X_a^d$ and two variable superpotentials $W_{ab}=X_a^d+X_b^d$. 
To each summand of the form $W_a$, we associate a fractional brane $\om_{l_a}$ while to each 
summand $W_{ab}$, we associate a rank one factorization $\op_{\eta_{ab}}$. 
Using these building blocks, we can construct new objects by taking tensor products. 
More precisely, let us decompose $W(X)$ as
\eqn\decomp{ 
W = W_{01}+W_{23}+\ldots +W_{2m,2m+1} + W_{2m+2}+\ldots + W_n}
for some $m< n/2-1$. 
Then we obtain an object 
\eqn\newfractA{ 
\oa_{\eta_{01}, \ldots, \eta_{2m,2m+1}; l_{2m+2},\ldots, l_{n}}=
\op_{\eta_{01}} \otimes \op_{\eta_{23}} \otimes \ldots \otimes \op_{\eta_{2m,2m+1}} \otimes 
\om_{l_{2m+2}} \otimes \ldots\otimes \om_{l_n}}
in the category $\CC_W$. Objects in the orbifold category can be obtained by making 
this construction equivariant with respect to the $\IZ/d$ action, 
as explained in the previous section. In this case it turns out that the $\IZ/d$ action on  
$\oa_{\eta_{01}, \ldots, \eta_{2m,2m+1}; l_{2m+2},\ldots, l_{n}}$ is completely 
determined by a single integer $\mu\in \{0,\ldots, d-1\}$.

For geometric applications, we need to determine the morphism spaces 
$$H^i(\om_{l_0,\ldots,l_n}, \oa_{\eta_{01}, \ldots, \eta_{2m,2m+1}; l_{2m+2},\ldots, l_{n}})$$
between the fractional branes and the new objects \newfractA.\ 
These spaces can be determined inductively, as explained in section 3.5. 
Following the steps detailed between \fractC\ and \fractE\ we first find 
\eqn\newfractB{ \eqalign{
& H^k\left(\om_{l_0,\ldots,l_3,\mu},\oa_{\eta_{01},\eta_{23},\mu'}\right) 
= \cr
& \mathop{\bigoplus_{i,j=0,1}}_{i-j\equiv k(2)} 
\bigoplus_{\alpha,\beta=0}^d H^i\left(\om_{l_0,l_1,\alpha},\op_{\eta_{01},0}\right)
\otimes H^j\left(\om_{l_2,l_3,\beta},\op_{\eta_{23},0}\right)\delta_{\mu'-\mu-(\alpha+\beta)}
\cr}}
for a single tensor product. This formula must be iterated each time we add an extra factor, 
which can be either $\op_{\eta_{a,a+1}}$ or $\om_{l_b}$ for some $a,b$ as in equation \fractI.\ 
Applying this algorithm, one can show that the intersection matrix can be obtained by multiplying 
the individual intersection matrices of the building blocks. From now on, 
we will restrict ourselves to objects with $l_a=1$ and denote the branes obtained by tensoring 
$k$ such objects by $\om^{k}$. The fractional branes will be denoted by 
$\om^{n+1}_{\mu}$ and the new objects \newfractA\ will 
be denoted by $\oa^{(m)}_{\mu}=\left(\op^{m+1}\otimes\om^{n-2m-1}\right)_{\mu}$. 
Then, using \rkoneCC\ we find
\eqn\newfractBA{ 
\chi\left(\om^{n+1}_{\mu},\oa^{(m)}_{\mu'}\right)= (G^{-2}-G^{-1})^{m+1}(\II-G^{-1})^{n-2m-1}.}
independently of the values of $\eta_{a,a+1}$ in the tensor product \newfractA.\ 
We will make use of this result in order to find 
the geometric interpretation of these objects in the large radius limit 
of the linear sigma model associated to \homsupB.\ 

\newsec{Geometric Interpretation in Linear Sigma Models} 

The Landau-Ginzburg orbifolds considered in this section are special points in
the SCFT moduli space associated to gauged linear sigma models. More
specifically, the linear sigma model in question has a single $U(1)$ gauge
field and $n+2$ chiral multiplets $X_0,\ldots, X_n, \Phi$ with charges
$(1,\ldots,1, -d)$. For $d=n+1$ the $U(1)_R$ symmetry of the linear sigma model
is anomaly free, and the infrared limit of the theory consists of a moduli
space of SCFT's parametrized by complexified FI terms.
Typically, this moduli space contains certain special points, where the SCFT
exhibits a special behavior. The special points of interest here are the Gepner
point, where the SCFT admits the LG orbifold description, and the large radius
point, where we have a nonlinear sigma model realization. In the last case, the
target space of the nonlinear sigma model is given by the Fermat hypersurface
$S$
\eqn\fermatA{ 
X_0^d+\ldots+X^d_n=0}
in $\IP^{n}$. For $d=n+1$, this is a Calabi-Yau variety.
These two limiting points are also called the Landau-Ginzburg and geometric phase respectively, 
although they are not separated by a sharp phase transition.   
In the previous sections we have discussed the construction of D-branes in the
Landau-Ginzburg phase from an algebraic point of view. We have recovered the
expected rational boundary states and also obtained new branes which do not
seem to have a boundary state realization. On the other hand, D-branes also
have a fairly explicit description in the geometric phase, where they should be
thought of as complexes of coherent sheaves, or more precisely, objects in the bounded
derived category $D^b(S)$ \refs{\K,\D,\AL}. 
Moreover, it is known that the Landau-Ginzburg fractional branes
can be analytically continued as BPS states to the large radius limit point,
and reinterpreted as holomorphic bundles on $S$. In particular, the fractional branes 
$\om^{n+1}_{\mu}$ correspond to the one term complexes $\underline{\Omega^{\mu}(\mu)}[\mu]$
$\mu=0,\ldots,d-1$ \refs{\BDLR,\DFRii}.
Here $\Omega^{\mu}$ denotes the bundle of holomorhic $\mu$-forms on the ambient projective space, 
and
$\Omega^{\mu}(\mu)=\Omega^{\mu}\otimes \CO(\mu)$. 
The notation $\underline{\Omega^{\mu}(\mu)}$ means
that $\Omega^{\mu}(\mu)$ should be thought of as a one term complex
concentrated in degree $0$, and $[\mu]$ denotes the shift functor of $D^b(S)$. 
Therefore $\underline{\Omega^{\mu}(\mu)}[\mu]$ represents a one term complex concentrated in 
degree $-\mu$. Note that all these bundles are
restricted to $S$ from the ambient projective space. 
For simplicity, we will denote these objects by $\Omega_\mu$. 

In the following we would like to address a similar question for the new 
fractional branes $\oa_{\mu}^{(m)}$ constructed in the previous section. 
For concreteness, we will consider the orbit $m=0$ for $n=4$, although similar methods 
can be applied to 
any values of $m,n$. The large radius hypersurface $S$ is the Fermat quintic in $\IP^4$. 

\subsec{Intersection Numbers and Topological Charges}

The objects $A^{(0)}_\mu$ correspond by analytic continuation to certain objects $\CF_\mu$ 
in the derived category $D^b(S)$. Our goal is to identify these objects using the algebraic 
structure 
developed so far at the Landau-Ginzburg point. 
A first useful observation is that 
marginal closed string perturbations preserve D-brane intersection numbers. 
This data suffices to determine the topological invariants of the objects $\CF_\mu$.
Then we will determine the actual objects (up to isomorphism) using the Landau-Ginzburg category structure. 

In the geometric phase, the intersection number between two D-branes represented by objects 
$\CF$ and $\CF'$ in the derived category is given by the alternating sum
\eqn\geomAA{
\chi(\CF,\CF')=\sum_{\nu\in\IZ}(-1)^{\nu}\hbox{dim}(\hbox{Hom}_{D^b(S)}(\CF,\CF'[\nu])).}
If $\CF$ is a one term complex consisting of a locally free sheaf $F$ in degree $0$, we have
\eqn\geomB{\chi(\CF[\mu],\CF') = (-1)^{\mu}\int_S 
\ch(F^\vee)\ch(\CF')\hbox{Td}(S),} 
where $F^\vee$ is the dual of $F$ and 
$\hbox{Td}(S)$ denotes the Todd class of $S$. Equation \newfractB\ predicts  
\eqn\geomC{ 
\chi\left(\Omega_\mu, \CF_{\mu'}\right) 
= -\left[(\II-G)^{4}\right]_{\mu,\mu'}.}
This yields a system of equations in the Chern characters of the unknown objects
$\CF_{\mu}$, which determines them uniquely. Of course, this is not enough data 
for identifying $\CF_{\mu}$ as objects 
in the derived category. We will later show how to determine the objects up to 
isomorphisms exploiting the algebraic structure. 

The even cohomology of the quintic hypersurface $S$ is generated by $(1,H,l,w)$ where $H$ is the 
class of a hyperplane section, $l$ is the class of a rational curve on $S$, and $w$ is 
the class of a point. The intersection ring is determined by the following relations
\eqn\intring{ 
H^2=5l,\quad Hl=1,\quad H^3=5w\,.}
A straightforward computation based on \geomB\ and \geomC\ yields the following Chern characters
\eqn\geomE{
\matrix {  & \ch_0 & \ch_1 & \ch_2 & \ch_3 \cr
\CF_0 & 1 & 0 & 0 & - w \cr
\CF_1 & -3 & H & {5\over 2}l & -{1\over 6} w \cr
\CF_2 & 3 & -2H & 0 & {7\over 3} w \cr
\CF_3 & -1 & H & -{5\over 2} l & -{1\over 6} w \cr
\CF_4 & 0 & 0 & 0 & -w \cr}\quad .}
A first check of this result is the integrality of the associated Chern classes:
an easy computation 
gives 
\eqn\geomF{
\matrix{ & c_0 & c_1 & c_2 & c_3 \cr
\CF_0\hfill & 1 & 0 & 0 & -2w \cr
\CF_1[1] & 3 & -H & 5l & -3w \cr
\CF_2 \hfill & 3 & -2H & 10l & -2w \cr
\CF_3[1] & 1 & -H & 0 & 2w \cr
\CF_4\hfill  & 0 & 0 & 0 & -2w \cr}\quad ,}
which are indeed integral. 

In principle, we would like to identify the objects $\CF_\mu$ in $D^b(S)$. 
Note that the Chern character of $\CF_4$ 
is equal to the Chern character of a skyscraper sheaf of length $1$ supported at a point $P$ 
on $S$ and also shifted by an odd integer. Since the shift functor is an automorphism of $D^b(S)$
we can choose this shift to be $-1$ without loss of generality.  
This strongly suggests that $\CF_4$ is a anti-D0-brane on the quintic. 
For the moment, we will conjecture that to be true. 

If that is the case, 
$\CF_\mu$, $\mu =0, \ldots, 4$ must form the orbit of the anti-D0-brane 
under monodromy transformations about the Landau-Ginzburg point. 
We will first test this conjecture by computing the 
Chern characters of the derived objects generated by ${\underline \CO}_P[-1]$ under $\IZ/5$ monodromy 
transformations. We will show below that this reproduces the table \geomE,\ which is strong evidence 
for our conjecture. Then in the next subsection we will show that $\CF_4$ is indeed a
anti-D0-brane supported at the 
point \hbox{$P=\{X_0-\eta X_1=X_2=X_3=X_4=0\}$} on the quintic using algebraic techniques. 
An interesting question is if one can give a physical proof of this conjecture based on RG 
flow in linear sigma models with boundary using the results of \refs{\HM,\GJS,\HKLM,\DJP}. 

\subsec{Landau-Ginzburg Monodromy} 

The monodromy about the Landau-Ginzburg point is generated by an 
autoequivalence ${\bf M}_{LG}$ of the 
derived category $D(S)$ which can be described as a Fourier-Mukai functor with kernel
\refs{\RPHi,\RPHii,\DJP} 
\eqn\LGmonA{ 
{\cal K}_{LG} = \hbox{Cone}\left({\bf L}\pi_1^*{\underline \CO} \otimes^{\bf L} {\bf L}\pi_2^* 
{\underline \CO}(1)\, \ra \,
{\underline {\CO_{\Delta}}}\otimes ^{\bf L} {\bf L}\pi_2^*{\underline \CO}(1)\right).}
The notation is standard. For any coherent sheaf $F$, we denote by ${\underline F}$ the one term 
complex determined by $F$ in degree zero. The maps $\pi_{1,2}:S\times S \ra S$ are canonical projections,
and $\CO(1)$ is the line bundle on $S$ obtained by restricting the hyperplane line bundle on $\IP^4$.  
$\CO_{\Delta}$ is the structure sheaf of the diagonal $\Delta\subset S\times S$. 
For any object $\CF$ in $D(S)$, the monodromy action is given by 
\eqn\LGmonB{ 
{\bf M}_{LG}(\CF) = {\bf R}\pi_{2*} \left({\bf L}\pi_1^*\CF \otimes^{\bf L} {\cal K}_{LG} \right).}
Our goal is to compute the monodromy orbit of ${\underline {\CO_P}}[1]$, where $\CO_P$ is the structure 
sheaf of a point $P$ on the quintic. The monodromy action \LGmonB\ admits an alternative description as a
twist functor \refs{\ST,\AHK}. For any object $\CF$ of $D(S)$, consider the  complex 
\eqn\LGmonC{ 
hom(\CO,\CF(1))\otimes \CO 
= \bigoplus_{k\in \IZ} \hbox{Hom}_{D^b(S)}(\CO,\CF(1)[k])\otimes_{\IC}{\underline \CO}[-k]
= \bigoplus_{k\in \IZ} {\underline \CO}[-k]^{\oplus b_k}}
where $b_k = \hbox{dim} \left(\hbox{Hom}_{D^b(S)}(\CO, \CF(1)[k])\right)$. 
Employing the techniques of \ST\ (see Lemma 3.2 and also \AHK),\ 
one can show that the Fourier-Mukai functor ${\bf M}_{LG}$ 
is isomorphic to the twist functor 
\eqn\LGmonD{ 
\CF \ra \hbox{Cone}\left(hom(\CO,\CF(1))\otimes \CO {\buildrel {ev}\over \longrightarrow } \CF(1)\right)}
where $ev$ is the evaluation map. 
Using this result, and the Riemann-Roch formula, we can easily compute 
\eqn\LGmonE{ \eqalign{
\hbox{ch}\left({\bf M}_{LG}(\CF)\right)) & = \ch(\CF(1)) -\sum_{k\in \IZ} (-1)^k b_k \cr
& = \ch (\CF(1))-\int_S \ch(\CF(1)) \hbox{Td}(S)\,.\cr}}
Now we can check by straightforward computations that 
\eqn\LGmonF{ 
\ch({\bf M}_{LG}^\mu 
({\underline {\CO_P}}[-1])) = \ch(\CF_{\mu +4})\,,}
as promised above.
In order to complete the picture, we will show below that $\CF_4$ is in fact isomorphic to 
the anti-D0-brane ${\underline {\CO_P}}[-1]$ using the category structure at the LG point. 

\subsec{The $D0$-brane at The Landau-Ginzburg Point}

The main idea is very simple. Let $j:S\ra \IP^4$ denote the embedding of $S$ into the projective 
space. Given a derived object $\CF$ on $S$, the pushforward ${\bf R}j_* \CF$ is obtained by extending the 
terms and the maps of $\CF$ by zero to $\IP^4$. Therefore, in order to determine $\CF$, it suffices to 
determine ${\bf R}j_* \CF$ in $D^b(\IP^4)$. 
This is a much simpler problem since $D^b(\IP^4)$ admits a pure 
algebraic description via Beilinson correspondence \refs{\B}. 
We will review some aspects below 
following \refs{\Bo}. 
 
According to Beilinson's theorem, 
$D^b(\IP^4)$ is generated by the exceptional collection $(\Omega^\mu(\mu))$, $\mu=0,\ldots, 4$.
This implies that there is an equivalence of categories between $D^b(\IP^4)$
and the derived category of 
$\CA$-modules, $D^b({\bf mod}-\CA)$, where $\CA$ is the endomorphism algebra 
\eqn\DOA{\CA=\hbox{End}(E),\qquad E=\oplus_{\mu=0}^4 \Omega^\mu(\mu).}
It is a standard result that $\CA$ is the path algebra of a finite ordered quiver with relations. 
For any projective space $\IP(V^\vee)$, with $V$ a complex vector space, we have 
\eqn\DOB{ 
\hbox{Hom}(\Omega^\mu(\mu), \Omega^\nu(\nu)) \simeq \left\{
\matrix{\Lambda^{\mu-\nu}(V),\qquad \hfill & \hbox{if}\ \mu\geq \nu\hfill\cr
0,\qquad\hfill & \hbox{otherwise .}\cr}\right.} 
The algebra structure of $\CA$ is determined by exterior multiplication. 

\ifig\endalg{The endomorphism algebra of $E$ as a quiver algebra.}
{\epsfxsize2.5in\epsfbox{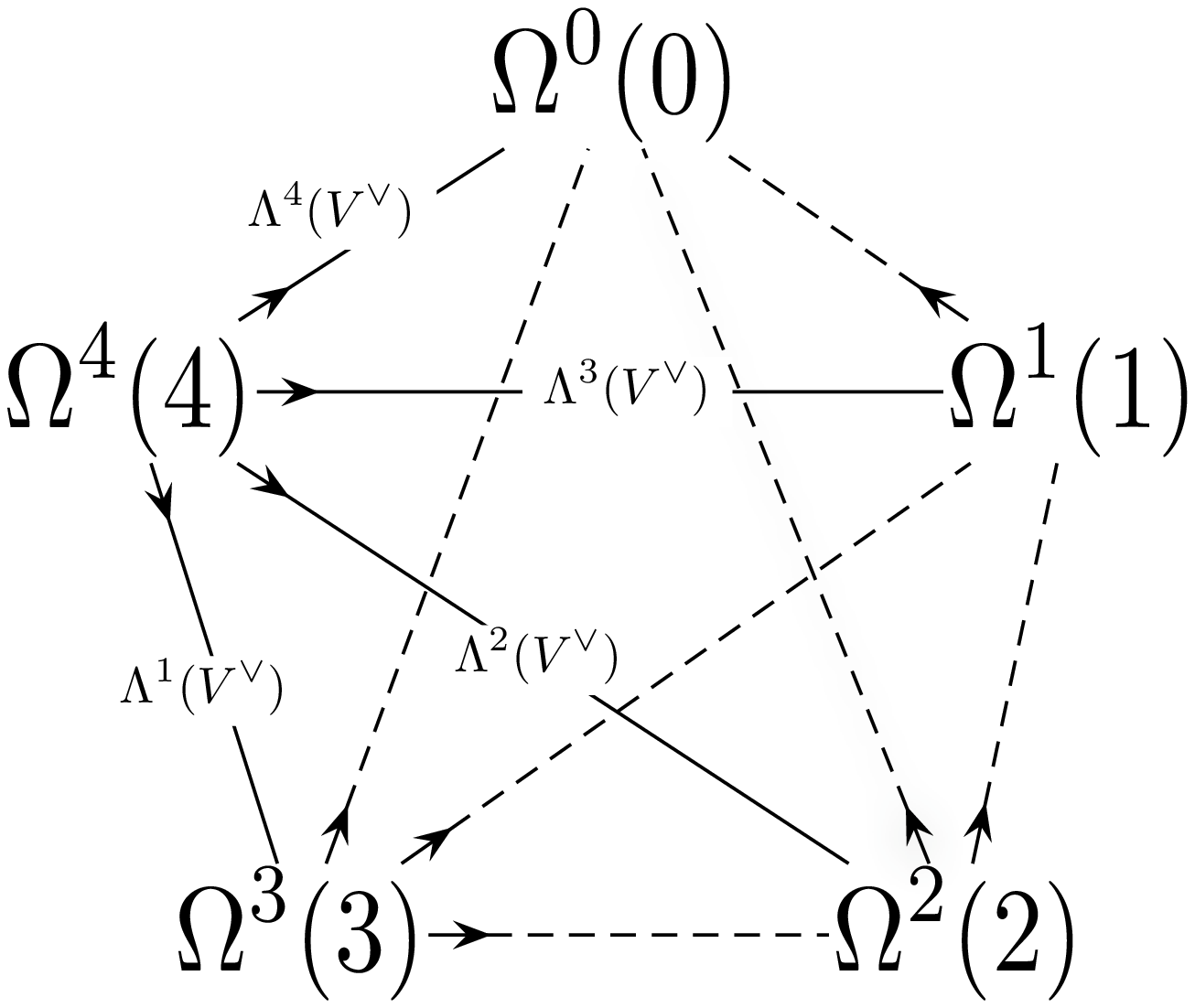}}

\noindent 
We can give a more geometric description of this algebra as follows. Note that we can identify 
the morphism spaces $\hbox{Hom}(\Omega^\mu(\mu), \Omega^{\mu-1}(\mu-1))$ to $H^0(T(-1))$, where 
$T$ is the holomorphic tangent bundle to $\IP^4$. Any section $v\in H^0(T(-1))$ determines a 
global morphism $\Omega^\mu(\mu)\ra \Omega^{\mu-1}(\mu-1)$ using the canonical pairing 
\eqn\pairing{ 
\Omega^\mu(\mu) \otimes T(-1) \ra \Omega^{\mu-1}(\mu-1).} 
One can check using the Euler exact sequence 
\eqn\euler{ 
0\ra \CO(-1) \ra \CO^{\oplus 5} \ra T(-1)\ra 0} 
that $H^0(T(-1))\simeq \IC^5$, hence we obtain all morphisms this way. 
Moreover, a basis of unit vectors in $H^0(\CO^{\oplus 5})\simeq \IC^5$ determines a 
basis of $H^0(T(-1))$. We will fix such a basis $\{v_a\}$, $a=0,\ldots 4$ from now on. 
This gives rise to a basis $v_a(\mu, \mu-1)$ in the space of morphisms 
$\hbox{Hom}(\Omega^\mu(\mu), \Omega^{\mu-1}(\mu-1))$ which is a system of generators 
of $\CA$. 

The equivalence of categories mentioned above associates to an object $\CF$ the
complex 
\eqn\DOC{ 
{\bf R}\hbox{Hom}({\underline E},\CF)= \bigoplus_{k\in \IZ} \hbox{Hom}({\underline E},\CF[k])[-k]}
of right $\CA$-modules with trivial differential. In particular this means that 
any object $\CF$ is uniquely determined 
(up to isomorphism) by the complex ${\bf R}\hbox{Hom}({\underline E},\CF)$. 

Our goal is to make use of this correspondence in order to find the derived object $\CF_4$ associated 
to the Landau-Ginzburg D-brane $\oa^{(0)}_4$. We will proceed in two steps. 

$i)$ We know that the fractional branes $\om^{5}_\mu$ correspond under 
closed string marginal deformations 
to ${\underline {\Omega^\mu(\mu)}}[\mu]$. On general grounds, these deformations
should preserve the category 
structure, hence we expect the endomorphism algebra 
\eqn\DOCA{{\overline \CA}=\hbox{End}^0\left({\overline E}\right),\qquad 
{\overline E} =  \bigoplus_{\mu=0}^4 \om^{5}_\mu[-\mu]}
to be isomorphic to $\CA$. As a first step, we will construct an explicit isomorphism $\phi:
{\overline \CA} \ra \CA$.  

$ii)$ 
At the next stage, we will determine 
the ${\overline \CA}$-module structure of the morphism space 
\eqn\morspace{
{\bf R}\hbox{Hom}({\overline E}, \oa^{(0)}_4)= \bigoplus_{k=0,1}H^k(\oe, \oa^{(0)}_4)} 
and compare it to the $\CA$-module structure of 
${\bf R}\hbox{Hom}({\underline E}, {\underline {\CO_{\IP^4,P}}}[1])$, where $P$ is the point 
$P=\{X_0-\eta X_1=X_2=X_3=X_4=0\}$ on $\IP^4$. $\CO_{\IP^4,P}$ is the structure sheaf of $P$
on $\IP^4$ which is the pushforward of the structure sheaf $\CO_{S,P}$ of $P$ on the quintic. 
More precisely, we will show that both complexes ${\bf R}\hbox{Hom}({\overline E}, \oa^{(0)}_4)$ 
and ${\bf R}\hbox{Hom}({\underline E}, {\underline {\CO_{\IP^4,P}}}[1])$ 
are concentrated in degree 
zero, and there is a linear isomorphism 
\eqn\DOCB{ 
\psi:  
{\bf R}\hbox{Hom}({\overline E}, \oa^{(0)}_4)\ra 
{\bf R}\hbox{Hom}({\underline E}, {\underline {\CO_{\IP^4,P}}}[1])}
so that $\psi({\overline a}t) = \phi({\overline a})\psi(t)$ for any ${\overline a}\in {\overline 
\CA}$, 
$t\in  {\bf R}\hbox{Hom}({\overline E}, \oa^{(0)}_4)$. 
However, ${\bf R}\hbox{Hom}({\overline E}, \oa^{(0)}_4)$ is isomorphic to the morphism complex 
${\bf R}\hbox{Hom}({\bf L}j^*{\underline E}, \CF_4)$ 
since the category structure must be preserved by 
closed string deformations. Furthermore, by adjunction, we have 
\eqn\adjunct{
{\bf R}\hbox{Hom}({\bf L}j^*{\underline E}, \CF_4) \simeq 
{\bf R }\hbox{Hom}({\underline E}, {\bf R}j_*\CF_4)\,.} 
It then follows from Beilinson's correspondence that 
${\bf R}j_*\CF_4$ is isomorphic to ${\underline {\CO_{\IP^4,P}}}[1]$, and we can conclude 
that $\CF_4$ is isomorphic to ${\underline {\CO}_{S,P}}[1]$.  

\subsec{Endomorphism Algebra}

In order to construct the isomorphism $\phi:{\overline \CA}\ra \CA$ 
let us first consider morphisms 
between fractional branes in some detail. These morphisms can be writen
as tensor products of one variable morphisms using the iterative algorithm developed in section 3.5, equations \fractD-\fractE.\ 

First consider morphisms between adjacent pairs $\om^5_\mu\,, \om^5_{\mu-1}$. 
Applying the rules of section 3.5 and the results of section 3.2, we find that there are 
no bosonic nor fermionic morphisms between $\om^5_{\mu-1}, \om^5_\mu$, and 
\eqn\endalgAA{\eqalign{ 
& H^0(\om^5_\mu, \om^5_{\mu-1})=0,\qquad H^1(\om^5_\mu,\om^5_{\mu-1})\simeq \IC^5, \qquad 
\mu=1,\ldots,5\,.\cr}}
An explicit basis 
$T_a(\mu,\mu-1)$, $a=0,\ldots,4$ of fermionic morphisms between $\om^5_\mu$ and $\om^5_{\mu-1}$ 
can be obtained by tensoring a one variable fermionic morphism 
associated to the superpotential $W_a(X_a)=X_a^5$ by four bosonic morphisms 
associated to the remaining variables. The later are all proportional to the identity. 
The result is most 
conveniently expressed in terms of free fermion operators as shown in section 2, equation 
2.8:
\eqn\endalgA{ 
T_a(\mu,\mu-1) = (-X_a^3\,\pib_a+\pi_a),\ldots, a=0,\ldots,4\,.}
Note that this expression is in fact independent of $\mu$ by $\IZ/5$ cyclic symmetry. 
However we have to use the notation 
$T_a(\mu, \mu-1)$ because different values of $\mu$ correspond to distinct generators of 
${\overline \CA}$. 

The other morphism spaces $H^i(\om^5_\mu, \om^5_\nu)$
can be similarly determined by taking tensor products 
of more one variable fermionic morphisms. 
Each such morphism contributes $-1$ to the orbifold charge. Therefore, by imposing equivariance, 
one finds that a morphism between objects with $\IZ_d$ weights $(\mu, \nu)$ must contain $\mu-\nu$ 
factors. A straightforward analysis shows that there are no bosonic nor fermionic morphisms 
between $\om^5_\mu$ and $\om^5_\nu$ if $\mu <\nu$. If $\mu\geq \nu$ we have 
$H^{\mu-\nu+1}(\om^5_\mu, 
\om^5_\nu)=0$, and $H^{\mu-\nu}(\om^5_\mu, \om^5_\nu)$ is generated by products of the form 
$T_{a_{\mu-\nu}}(\nu+1,\nu)\ldots T_{a_1}(\mu,\mu-1)\,$. 
For concreteness we represent below the morphism spaces from $\om^5_4$ to all other 
fractional branes in the orbit. 

\ifig\quintic{Morphisms from $\om^5_{4}$ to the other fractional branes in the orbit.}
{\epsfxsize2in\epsfbox{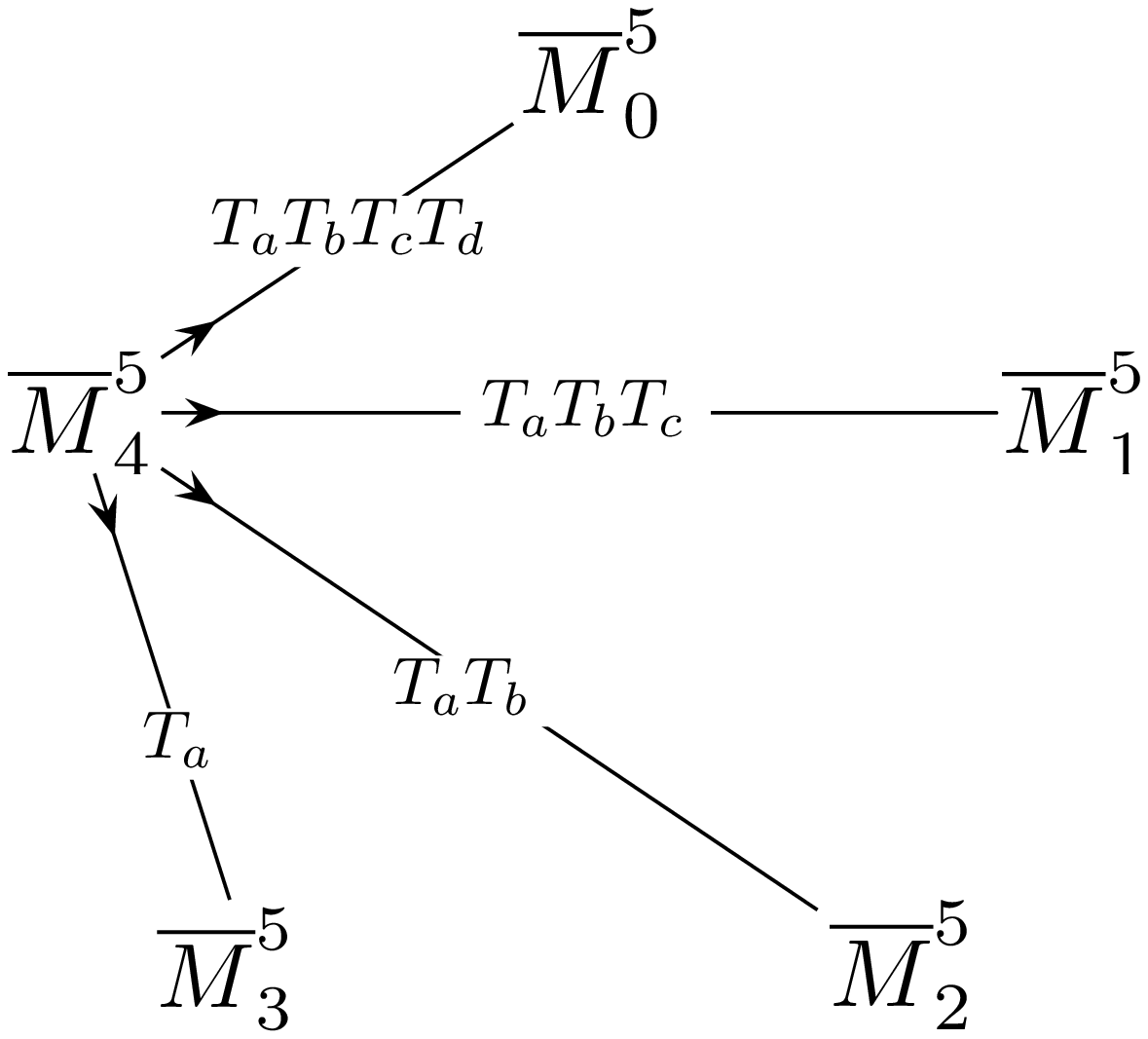}}

\noindent
Note that the resulting morphisms are bosonic for even number of factors and fermionic for 
odd number of factors.
Morphisms constructed this way are not all independent. 
Using the canonical anticommutation relations 
\anticom\ it follows that the $T_a$ satisfy the algebra
\eqn\Talgebra{
T_a(\mu,\mu-1)T_b(\mu-1,\mu-2) + T_b(\mu,\mu-1) T_a(\mu-1,\mu-2) =0\,.}
In particular, $T_a(\mu,\mu-1)T_a(\mu-1,\mu-2) = 0$. 
These are of course cohomology relations; at cochain level the right hand side is 
an exact cochain in the complex \lgA.\ 
Then it follows that the number of independent morphisms between any pair of objects 
equals the number of antisymmetric combinations of products of $T_a(\mu,\mu-1)$. 
Therefore, we have  
\eqn\TalgebraB{H^{\mu-\nu}(\om_{\mu}^{5},\om_{\nu}^5)\simeq \Lambda^{\mu-\nu}(U^{(5)}),\quad
\hbox{for}\ \mu\geq \nu\,,}
where $U^{(5)}$ is the five dimensional complex vector space spanned by 
$\{T_a\}$, $a=0,\ldots, 4\,$. 

It is worth noting that this construction can be applied without essential 
changes to more general situations. 
For example, suppose we want to determine the morphism spaces between fractional branes 
associated to a 
$(n+1)$-variable superpotential of degree $d$, where $n$ and $d$ are arbitrary. One finds a very similar 
structure, that is  $H^{\mu-\nu}(\om^{n+1}_\mu, \om^{n+1}_\nu)\simeq \Lambda^{\mu-\nu} (U^{(n+1)})$ 
for $\mu\geq \nu$, where $U^{(n+1)}$ is a $(n+1)$-dimensional complex vector space spanned by 
$\{T_a\}$, $a=0,\ldots, n+1$. If $\mu<\nu$, the morphism spaces are empty.
The generators are given again by products of fermionic operators $T_a(\mu,\mu-1)$, 
$a=0,\ldots, n\,$, $\mu=0,\ldots,d-1\,$. The formula \endalgA\ becomes 
\eqn\endalgAX{ 
T_a(\mu,\mu-1) = -X_a^{d-2} \pib_a+\pi_a\,,}
except that now we have to pick a different representation of the Clifford algebra. The relations 
\Talgebra\ remain unchanged. We will find this remark very useful in the next subsection. 

Let us now define the object 
\eqn\obj{
\oe=\bigoplus_{\mu}\oe_{\mu}}
where $\oe_{\mu}=\om^5_{\mu}$ for $\mu=0,1,2$ and $\oe_{\mu}=\om_{\mu}^5[1]$ 
for $\mu=1,3$. 
From the definition of the shifted object \shift,\ we derive the following 
rules: for each even (odd) morphism $\Phi$ from $\op$ to $\oq$,
there is an odd (even) morphism $\Phi\cdot J$ from the shifted object
$\op[1]$ to $\oq$ and an odd (even) morphism $J\cdot\Phi$ from $\op$ to 
$\oq[1]$. Here, $J$ is an odd operator such that
$J^2=1$. In a matrix representation where bosonic operators are diagonal
blocks and fermionic ones are off diagonal, $J$ has the form
$$
J=\left(\matrix{0 & \II \cr \II & 0}\right).
$$
Therefore, using \TalgebraB,\ we find  
\eqn\endalgB{
 \hbox{End}^0(\oe_{\mu},\oe_{\nu})\simeq \left\{\matrix{\Lambda^{\mu-\nu}
(U^{(5)}) & \mu \ge \nu \cr
0 &\quad\quad \hbox{otherwise .}} \right.}

\ifig\Emorphisms{Endomorphisms of the object $\oe$}{\epsfxsize2.2in\epsfbox{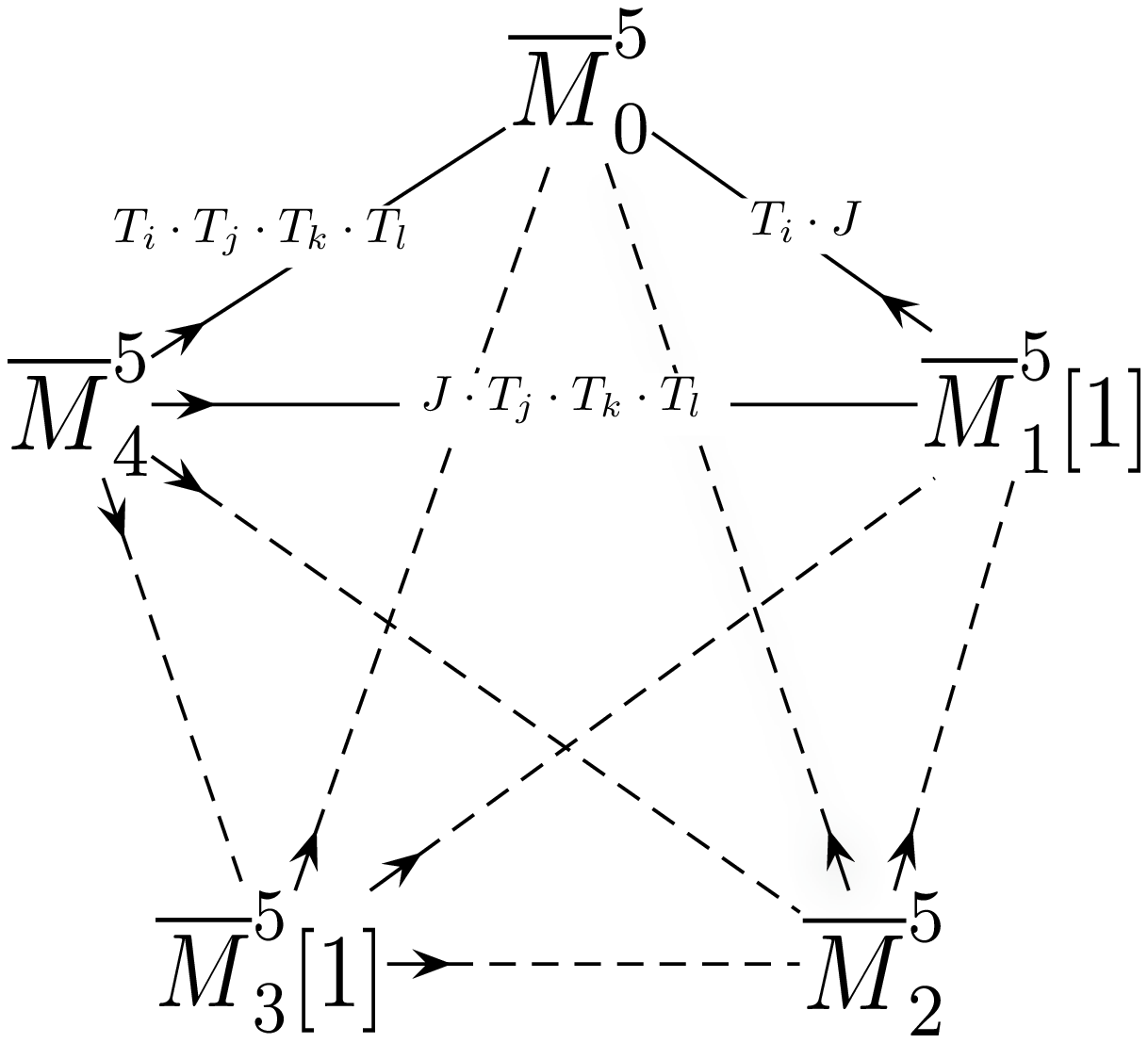}}

\noindent 
The algebra of bosonic endomorphisms of $\oe$ is determined by the basic relations 
\Talgebra\ and the additional relation $J^2=1$. One finds that ${\overline \CA}$ is generated by 
the elements $T_a(\mu, \mu-1)\cdot J$ and the algebra structure is given by exterior 
multiplication. 
A concrete example is represented in \Emorphisms.\ 

In conclusion, $\oa$ is indeed isomorphic to the geometric quiver algebra 
$\CA$ and an explicit isomorphic $\phi: {\overline \CA} \ra \CA$ is given by 
\eqn\endalgC{\phi(T_a(\mu,\mu-1)J) = 
v_a(\mu, \mu-1)\,.}

\subsec{Module Structure} 

The next step is to determine the ${\overline \CA}$-module structure of ${\bf R}\hbox{Hom}(\oe,\oa^{0}_4)$ 
and construct the isomorphism \DOCB.\ We first have to determine ${\bf R}\hbox{Hom}(\oe,\oa^{0}_4)$
as a vector space. The morphims from $\oe_\mu$ to $\oa^5_\nu$ can be constructed
again by taking tensor products as in the previous subsection. Recall that $\oa^(0)_\nu$ is obtained 
by tensoring a rank one factorization $\op_4$ of the two variable superpotential $W_{02}(X_0, X_1) = 
X_0^5+X_1^5$ coresponding to some fixed 
$\eta$, $\eta^5=-1$ by $l=1$ factorizations of the remaining one variable superpotentials, i.e. 
$\oa^0_\nu = \left(\op \otimes \om^3\right)_\nu$. 
The fractional branes $\om^5_\mu$ can be accordingly writen as tensor products of the form 
$\left(\om^2\otimes \om^3\right)_\mu$ corresponding to the variables $(X_0, X_1)$ and respectively 
$(X_2, X_3, X_4)$. 
Using the methods of section 3.5 (see the paragraph between \fractC\ and \fractD,)\ we can
easily show that the morphism spaces between $\om^5_\mu$ and $\oa^0_\nu$ are given by 
\eqn\moduleA{ 
H^k(\om^5_\mu, \oa^0_\nu) = \mathop{\bigoplus_{i,j=0,1}}_{i+j=k(2)} \bigoplus_{\mu',\nu'=0}^4 
H^i(\om^2_{\mu'},\op_0) \otimes H^j(\om^3_{\nu'}, \om^3_0)\delta_{\mu'+\nu' -(\mu-\nu)}.}
The morphism spaces between $\om^2_{\mu}$ and $\op_\nu$ have been determined 
in section 4.1, equation \rkoneCB.\ 
Up to multiplication by a nonzero complex number, we have 
one bosonic morphism $f_{\mu,\nu}$ if $\nu=\mu-2\ (5)$ and one fermionic morphism $t_{\mu,\nu}$ if 
$\nu=\mu-1\ (5)$. The morphism spaces between the three variable fractional branes can be determined 
by setting $n=2$ in the discussion above equation \endalgAX.\ 
We have 
$H^{\mu-\nu}(\om^3_\mu,
\om^3_\nu)\simeq \Lambda^{\mu-\nu}(U^{(3)})$ if $\mu\geq \nu$ and zero otherwise. 
Collecting these results, and taking into account the shifts by one, we find 
\eqn\moduleB{ 
H^0(\oe_\mu, {\oa^{(0)}_4}) \simeq 0,\qquad 
H^1(\oe_\mu, {\oa^{(0)}_4}) \simeq \Lambda^{\mu} (U^{(4)})\,,}
where $U^{(4)}$ is the four dimensional complex vector space spanned by 
$(f_{01}\otimes \II)J$ and \hbox{$(t_{04} \otimes T_2)J\,$,} $(t_{04}\otimes
T_3)J\,$, $(t_{04}\otimes T_4)J$.  

The ${\overline A}$-module structure is determined by composition with the 
morphisms \hbox{$T_a(\mu, \mu-1)J$} between fractional branes, which generate $\oa$. 
Let us consider for example multiplication by $H^0(\oe_1,\oe_0)$ as represented below. 

\ifig\EtoA{Module structure of ${\bf R}\hbox{Hom}(\oe, {\oa^{(0)}_4})$: multiplication by 
$H^0(\oe_1,\oe_0)$.}
{\epsfxsize2.0in\epsfbox{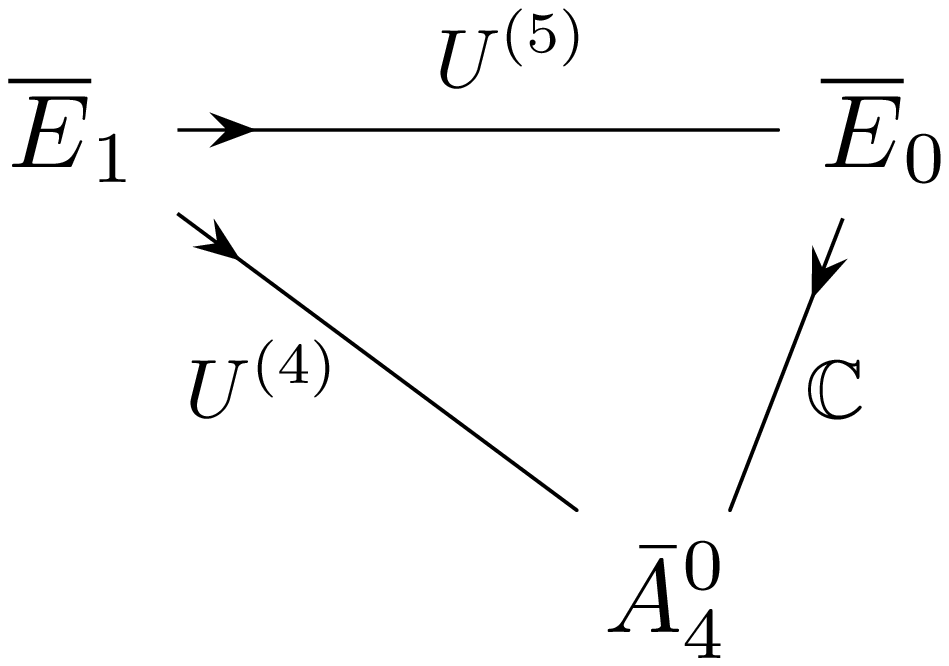}}

\noindent
We have established that  $H^1(\oe_0,{\oa^{(0)}_4})\simeq \IC$ and
$H^1(\oe_1,{\oa^{(0)}_4})\simeq U^{(4)}$. Next, 
our goal is to determine the composition laws of the morphisms in the above diagram. 
To this end, we have to write down the generators of all morphism spaces in terms 
of tensor products of elementary morphisms as in the previous subsection. 
Making the product structure explicit, we obtain the diagram represented in fig.8. 
 
\ifig\EtoP{Morphism generators.}
{\epsfxsize4.0in\epsfbox{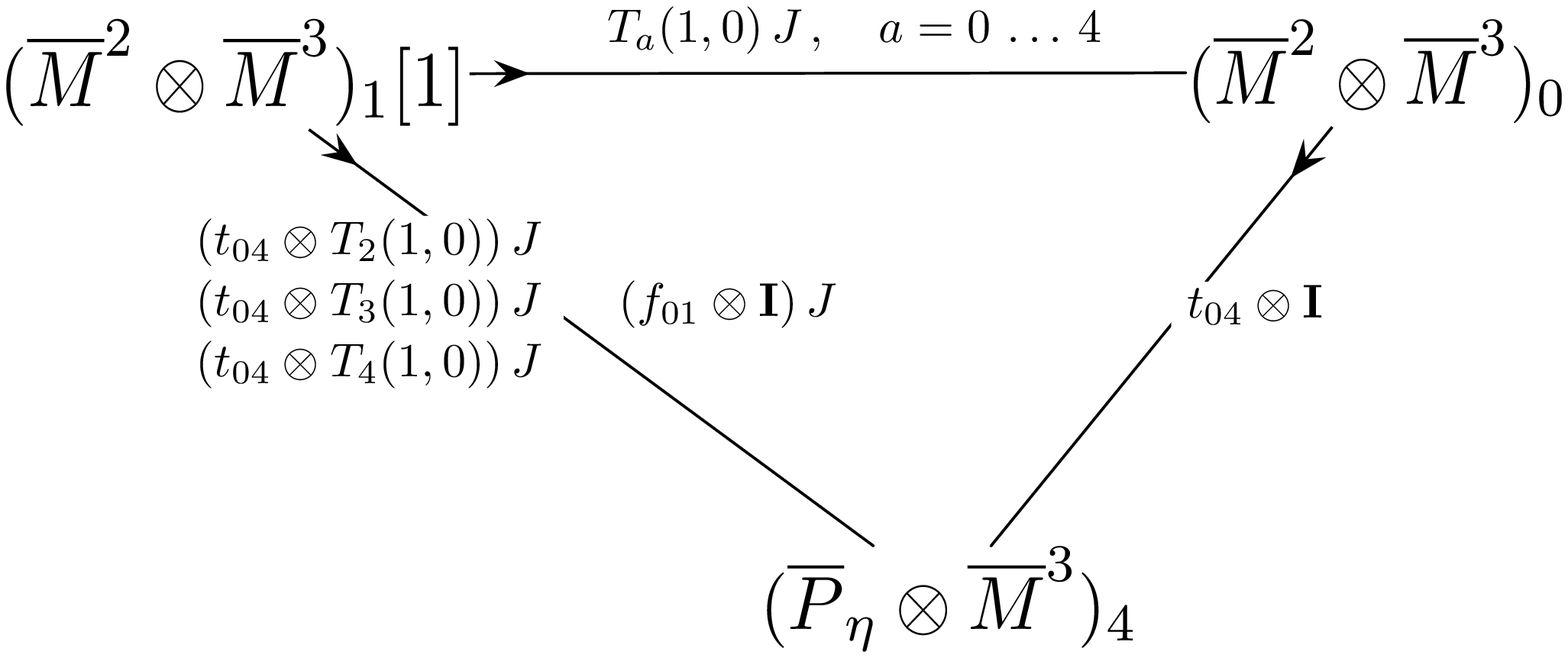}}

To explain the notation, recall that $f_{\mu,\nu}$ and $t_{\mu,\nu}$ are bosonic and respectively 
fermionic morphisms between two variable objects as explained in the paragraph above equation 
\moduleB.\ 
On the horizontal arrow, $T_a(1,0)J$ are the generators constructed in the previous subsection. 
On the left arrow, $T_{2,3,4}(1,0)$ denote similar fermionic morphisms between three variable 
objects $\om^3_1,\om^3_0$ introduced in the paragraph above \obj.\ 
We have used the same
notation in order to avoid unnecessary indices. The distinction should be clear from the context. 
In the next paragraph we will use the same conventions for $T_{1,2}(1,0)\,$, regarded as fermionic 
morphisms between two variable fractional branes. 

Since the tensor product is compatible with composition of morphism, we have the 
following straightforward composition laws 
\eqn\moduleBB{
(t_{04}\otimes I) (T_a(1,0)J) = (t_{04}\otimes T_a(1,0))J,\qquad a=2,3,4.}
In order to complete the picture, we have to determine the remaining products corresponding 
to $a=0,1$. Note that in this case, the fermionic morphisms $T_a(1,0)$, $a=0,1$ are obtained by
tensoring two variable fermionic morphisms $T_a(1,0)$ by bosonic morphisms in the remaining 
three variables, which are proportional with the identity. Since all morphisms involved in this 
computation are proportional to the identity in the last three variables, we are effectively left with a 
two variable problem. More precisely, it suffices to determine the products $t_{04}T_a(1,0)J$, $a=0,1$ 
for two variable morphisms. 
This can be done by the explicit computation presented in appendix B. We find the 
relations 
\eqn\moduleC{ 
t_{04}(T_1(1,0)J) = f_{01}J,\qquad 
t_{04}((\eta T_0(1,0) + T_1(1,0))J)=0\,.}

The multiplication table by the remaining generators $T_a(\mu,\mu-1)$ of ${\overline \CA}$ 
can be determined 
similarly, using the relations \Talgebra\ among the fermionic generators and the equations 
\moduleC.\ We will not present all the details here because they are quite tedious. 
It suffices to note 
that the linear combination $(\eta T_0(\mu,\mu-1) + T_1(\mu, \mu-1))J$ always 
annihilates $ H^1(\oe_{\mu-1},{\oa^{(0)}_4})$, i.e. 
\eqn\moduleD{
(\eta T_0(\mu,\mu-1) + T_1(\mu, \mu-1))J \cdot H^1(\oe_{\mu-1},{\oa^{(0)}_4}) = 0}
for $\mu=0,\ldots,4\,$. Moreover, one can check that  
\eqn\moduleE{ 
H^0(\oe_\mu,\oe_{\mu-1})/\langle (\eta T_0(\mu,\mu-1) + T_1(\mu, \mu-1))J \rangle
\simeq U^{(4)}} 
acts on $H^1(\oe_{\mu-1},{\oa^{(0)}_4})\simeq \Lambda^{\mu-1}(U^{(4)})$ 
by exterior multiplication. More precisely, the pairing 
\eqn\moduleF{
H^0(\oe_\mu,\oe_{\mu-1})/\langle (\eta T_0(\mu,\mu-1) + T_1(\mu, \mu-1))J\rangle 
\otimes H^1(\oe_{\mu-1},{\oa^{(0)}_4}) \,\ra \,H^1(\oe_\mu,{\oa^{(0)}_4})}
is isomorphic to the pairing 
\eqn\moduleG{ 
U^{(4)} \otimes \Lambda^{\mu-1}(U^{(4)}) \,\ra \,\Lambda^\mu(U^{(4)})\,.} 
This completes the description of ${\bf R}\hbox{Hom}(\oe,{\oa^{(0)}_4})$ as an 
${\overline \CA}$-module. 

In order to complete the picture, we have to construct the linear map \DOCB\ and prove 
the compatibility with the module structures. To this end we need a more detailed  
description of 
the $\CA$-module structure of 
${\bf R}\hbox{Hom}({\underline E}, {\underline{\CO_{\IP^4, P}}}[-1])$. 

According to the defining equation \DOC,\ 
the terms in this complex are derived morphism spaces of the form 
$\hbox{Hom}({\underline E},{\underline{\CO_{\IP^4, P}}}[k-1])\simeq \hbox{Ext}^{k-1}
({\underline E}, {\underline{\CO_{\IP^4, P}}})$. Since $E$ is locally free, the only nonzero 
term is 
\eqn\moduleH{\hbox{Ext}^0({\underline E},  {\underline{\CO_{\IP^4, P}}}) \simeq
E^\vee_P\,,}
where $E^\vee_P$ is the fiber of the dual bundle at $P$. We have 
\eqn\moduleI{ 
E^\vee_P = \bigoplus_{\mu=0}^5 (\Lambda^\mu T)(-\mu)_P = \bigoplus_{\mu=0}^5 \Lambda^\mu(T(-1)_P),}
hence 
\eqn\moduleIA{ 
{\bf R}\hbox{Hom}({\underline E}, {\underline{\CO_{\IP^4, P}}}[-1])\simeq 
 \bigoplus_{\mu=0}^5 \Lambda^\mu(T(-1)_P)[-1].}

In order to determine the $\CA$-module structure of this vector space, it suffices to determine 
the multiplication table by the generators $v_a(\mu,\mu-1)\in H^0(T(-1))$ introduced 
below \euler.\ These are twisted global holomorphic vector fields on $\IP^4$ associated to the 
unit vectors in $H^0(\IP^4, \CO_{\IP^4}^{\oplus 5})\simeq \IC^5$, as explained there. Let 
$v= \sum_{a=0}^4 \rho_a v_a(\mu,\mu-1)$ be an arbitrary linear combination of the generators. 
Let $P\in \IP^4$ be a point defined by intersecting four hyperplanes given by the homogeneous 
linear polynomials $L_1,\ldots,L_4$, that is $P=\{L_1=\ldots =L_4=0\}$. 

Since $T(-1)$ is a rank four bundle, a generic section $v$ as above is expected 
to vanish along a collection of points in $\IP^4$. 
From the Euler sequence it follows that the 
section $v$ vanishes at $P$ if and only if the point $(\rho_0,\ldots, \rho_4)\in \IC^5$ 
lies on the line through the origin defined by $P$. Therefore $v$ has a zero at $P$ 
if and only if 
\eqn\moduleJ{
L_1(\rho_0,\ldots,\rho_4) =\ldots = L_4(\rho_0,\ldots,\rho_4) =0\,.} 
This shows that any section $v$ vanishes precisely at one point $P_v\in \IP^4$ determined
by the equations 
\moduleJ.\ Conversely, for any $P$ there is a unique section $v_P\,$, up to 
multiplication by a 
nonzero constant, such that $v_P(P)=0\,$. 
In our case $P=\{X_0-\eta X_1=X_2=X_3=X_4=0\}$, hence $v_P = \eta\, v_0 + v_1\,$. 

Now we can easily determine the $\CA$-module structure of \moduleIA.\ For each pair 
$(\mu,\mu-1)$ there is a unique (up to scale) morphism $v_P(\mu,\mu-1) \in 
\hbox{Hom}(\Omega^\mu(\mu), \Omega^{\mu-1}(\mu-1)) \simeq H^0(T(-1))$ which annihilates 
$(\Lambda^{\mu-1}T)(-(\mu-1))_P$. 
Using the Euler sequence, it follows that the quotient 
$\hbox{Hom}(\Omega^\mu(\mu), \Omega^{\mu-1}(\mu-1))/\langle v_P(\mu,\mu-1)\rangle$ is 
isomorphic to $T(-1)_P$. 
Therefore the composition of morphisms determines a well defined pairing 
\eqn\moduleL{ 
T(-1)_P 
\otimes \Lambda^{\mu-1}(T(-1)_P)\ra \Lambda^\mu (T(-1)_P)}
which is dual to the canonical pairing \pairing\ 
$$\Omega^\mu(\mu) \otimes T(-1) \ra \Omega^{\mu-1}(\mu-1)$$
restricted to $P$. 
Therefore the pairing \moduleL\ is defined by exterior multiplication. 
This completes the description of the $\CA$-module structure of \moduleIA.\ 

Now we can collect all loose ends and complete the identification between Landau-Ginzburg and 
geometric data. In the Landau-Ginzburg category we found 
\eqn\moduleM{
{\bf R}\hbox{Hom}(\oe,{\oa^{(0)}_4}) \simeq \bigoplus_{\mu=0}^4 
\Lambda^\mu(U^{(4)})[-1]\,,}
where $U^{(4)}$ is a fixed four dimensional complex vector space isomorphic to 
$$H^0(\oe_\mu,\oe_{\mu-1})/\langle (\eta T_0(\mu,\mu-1) + T_1(\mu,\mu-1))J\rangle$$ 
for any $\mu=0,\ldots 4$. The ${\overline A}$-module structure is determined by exterior multiplication
as shown in equations \moduleF-\moduleG.\ This is the same structure as in the geometric 
situation. 
More precisely, the map $\phi$ in  \endalgC\ induces an isomorphism of vector spaces 
\eqn\moduleN{ 
{\bar \phi} : U^{(4)}\simeq H^0(\oe_\mu,\oe_{\mu-1})/\langle (\eta T_0(\mu,\mu-1) 
+ T_1(\mu,\mu-1))J\rangle \rightarrow  
H^0(T(-1))/\langle \eta v_0 + v_1 \rangle \simeq T(-1)_P.}
This can be extended by exterior multiplication and direct sums to a linear isomorphism 
\eqn\moduleP{
\psi : \bigoplus_{\mu=0}^4 \Lambda^\mu(U^{(4)})[-1]
\ra \bigoplus_{\mu=0}^5 \Lambda^\mu(T(-1)_P)[-1]}
which is clearly compatible with the module structures. 

\newsec{Composite Objects and Deformations} 

In this section we discuss several applications of our construction to deformations of D-branes 
at the Landau-Ginzburg point. This is not meant to be a complete and rigorous treatment of 
moduli problems in the Landau-Ginzburg category.
We will outline some preliminary results in two examples in 
order to illustrate the main ideas. We leave a more detailed approach for future work. 

\subsec{Composites of Fractional Branes} 

The first problem we would like to address here was formulated in \DGJT.\ As explained in 
the above section, 
we have five bosonic morphisms between two consecutive shifted fractional branes 
$\oe_\mu$, $\oe_{\mu-1}$. 
Then we can form composite objects by taking cones over these morphisms in the 
Landau-Ginzburg category. 
One of the main questions considered in \DGJT\ was counting the number of deformations of such 
composite objects, and finding equivalence relations between them. These questions 
may be quite difficult in the boundary state or the quiver gauge theory approach. 
Here we will address this issue from the 
algebraic 
point of view adopted in this paper. 

As a concrete example, we will consider two different composite objects 
which were conjectured to be
isomorphic branes in \DGJT.\ 
Using the quiver gauge theory approach, one can prove that these objects have the same 
moduli space, but not that they are indeed isomorphic. 
We will apply our formalism in order to prove this conjecture by 
constructing an explicit isomorphism between them. 

The first object is the cone 
$C_{10} = \hbox{Cone}\left(\oe_1\rra{\mathop{}_{ T_a(1,0)J}} \oe_0\right)$. 
In order to construct the second object, we have to take two successive cones. 
We first take the cone $C_{32}=\hbox{Cone}
\left(\oe_3\rra{\mathop{}_{T_a(3,2)J}} \oe_2\right)$.
Then we have an exact triangle 
\eqn\triangleAX{ 
\oe_3 \rra{\mathop{}_{T_a(3,2)J}} \oe_2 \ra C_{32}}
which yields the following long exact sequence of morphism groups 
\eqn\exseqA{\eqalign{
\cdots& \ra H^{-1}(\oe_4,\oe_2)\ra H^{-1}(\oe_4,\C_{32})
\ra H^0(\oe_4,\oe_3) \rra{\mathop{}_{T_a(3,2)J}}
H^0(\oe_4,\oe_2)\cr 
& \ra H^0(\oe_4,C_{32})\ra H^1(\oe_4, \oe_3)\ra \cdots }}
Since there are no fermionic morphisms between any pair $(\oe_\mu,\oe_\nu)$, 
this sequence reduces to 
\eqn\exseqB{ 
0\ra  H^{-1}(\oe_4,C_{32})
\ra H^0(\oe_4,\oe_3)  \rra{\mathop{}_{T_a(3,2)J}}
H^0(\oe_4,\oe_2) \ra H^0(\oe_4,C_{32})\ra 0}
where the middle map is induced by multiplication by $T_a(3,2)J$. 
Using the algebra structure of ${\overline \CA}$
determined above, it follows that $\hbox{Ker}(T_a(3,2)J)$ is one dimensional and
 is generated by 
$T_a(4,3)J$. Therefore we conclude that the space of fermionic morphisms 
$H^{-1}(\oe_4,C_{32})\simeq H^0(\oe_4[1], C_{32})$ is one dimensional. 
Let $\kappa$ denote a generator. 
Now consider the cone 
$C_{432} =\hbox{Cone}\left(\oe_4[1]{\buildrel\kappa\over \ra } C_{32}\right)$. 
The conjecture of \DGJT\ is that $C_{10}$ and $C_{432}[1]$ must be isomorphic in the D-brane 
category. This is required for agreement with geometric considerations in the large radius limit. 

Here we will show that this is true by constructing a morphism 
$\gamma : C_{10}\ra C_{432}[1]$ 
such that $\hbox{Cone}\left(C_{10} {\buildrel \gamma\over \ra} C_{432}[1]\right)$ 
is the trivial 
object in the Landau-Ginzburg orbifold category. 

A very useful observation is that cones commute with the tensor product defined in section 3.4. 
More 
precisely, suppose we have two pairs of objects $(\op,\op')$ and $(\oq,\oq')$ corresponding 
to two superpotentials $W_1, W_2$ as in that section. Suppose we are given two 
bosonic morphisms 
$\gamma:\op\ra \op'$ and $\delta : \oq\ra \oq'$. By taking the tensor product we obtain a bosonic 
morphism $\gamma\otimes \delta : \op\otimes \oq\,\ra\, \op'\otimes \oq'$. We claim that 
\eqn\coneprod{\hbox{Cone}\left(\op\otimes \oq 
\rra{\mathop{}_{\gamma\otimes \delta}}\op'\otimes 
\oq'\right) \simeq \hbox{Cone}\left(\op{\buildrel \gamma\over \ra} \oq\right) \otimes 
\hbox{Cone}\left(\oq{\buildrel \delta \over \ra } \oq'\right).}
The proof of this statement is a straightforward check using the definitions. 
We will not spell out the
details here. 

Now recall that the fractional branes $\oe_\mu$ are obtained by taking tensor products of one 
variable 
factorizations, and the morphisms $T_a(\mu,\mu-1)$ are similarly obtained by taking tensor products
by the identity.  
Using equation \coneprod,\ we can reduce the 
conjecture formulated above to a statement concerning  
one variable factorizations, which is much easier to prove. In the following we denote by 
$\om_{l,\mu}$, 
$l=1,\ldots, 4$, 
$\mu =0,\ldots, 4$ the rank one factorizations constructed in section 3.2 for the superpotential 
$W_a(X_a) = X_a^5$. With abuse of notation, we will use the same notation as above for cones in the 
orbifold category of $W_a$. 

The structure of exact triangles in a one variable category in the absence of an orbifold action 
has been determined in \O.\ In particular one can show that there are distinguished triangles 
 of the form 
\eqn\triangleB{ 
\om_{l}{\buildrel \phi_{lk}\over \ra}\om_{k} \ra \om_{k-l}}
where $\phi_{lk} :\om_{l}\ra \om_k$ is given by $(\phi_{lk})_1= x^{l-k},\ (\phi_{lk})_0=1$. 
If $k<l$, $\om_{k-l}$ is defined to be $\om_{l-k}[1]$. 
In the presence of an orbifold projection, the equivariant version of \triangleB\ is 
\eqn\triangleC{ 
\om_{l,\mu}{\buildrel \phi_{lk}\over \ra} \om_{k,l-k+\mu} \ra \om_{k-l,l-k+\mu}}
where the labels $\mu,\nu,\ldots \in \{0,1,\ldots, 4\}$ are defined mod 5. 
Moreover, we have the identifications $\om_{l,\mu}[1] \simeq \om_{-l,l+\mu}$ 
explained in section 
3.2. 

Now let us repeat the above cone construction in the one variable theory. We define 
$C_{10}=\hbox{Cone}\left(\om_{1,1}[1] \rra{\mathop{}_{T_a(1,0)}} \om_{1,0}\right)$. 
From \triangleC\ it follows that $C_{10} \simeq \om_{2,0}$. Next we construct 
$C_{32} =\hbox{Cone}\left(\om_{1,3}[1] \rra{\mathop{}_{T_a(3,2)}} \om_{1,2} \right)$
which is isomorphic to $\om_{2,2}$ by the same argument. Finally we take a second cone 
$C_{432}=\hbox{Cone} 
\left(\om_{1,4}[1] {\buildrel \kappa \over \ra} C_{32}\right)$. Using \triangleC\
again, we find $C_{432} \simeq \om_{3,2}$, hence $C_{432}[1]\simeq \om_{-3,0}$. 
To conclude, note that we have an exact triangle 
\eqn\triangleD{ 
\om_{2,0} \ra \om_{-3,0} \ra \om_{-5,0}}
where $\om_{-5,0}$ is the trivial object. Therefore we have indeed $C_{10}\simeq C_{432}[1]$. 
The same conclusion is valid for the five variable fractional branes by taking tensor products and 
using \coneprod.\

\subsec{$D0$-brane moduli} 

The second problem considered in this section is finding the moduli space of the object 
${\oa^{(0)}_4}$ which has been identified with a (anti) D0-brane on the quintic. Ideally one should 
be able to prove that this moduli space is isomorphic to the Fermat quintic\foot{In order to 
give a rigorous construction of the moduli space, one has to first specify a stability 
condition. In principle the moduli space may be different for different stability conditions. 
Therefore a more precise statement would be that the moduli space should be isomoprhic 
to the Fermat quintic in the presence of a suitable stability condition. 
We will not try to make this explicit here, although it is a very interesting subject 
for future work.} We will not give a 
rigorous proof here since moduli problems in abstract categories are complicated and not very well 
understood at this point. However we will take a first step in this direction by exhibiting a 
family of deformations of ${\oa^{(0)}_4}$ parameterized by the Fermat quintic. Although we cannot prove 
that this is the full moduli space in the D-brane category, this is certainly an important step 
forward. 

Let us consider a point $P$ on the Fermat quintic determined by the linear homogeneous equations 
$L_1=L_2=L_3=L_4=0$. According to Hilbert's Nullstellensatz, 
the condition that $P$ lie on the quintic $W=0$ is that $W^q$ belongs to the 
ideal $I=(L_1,L_2,L_3,L_4)\subset \IC[X_0,\ldots,X_4]$ for some $q>0$. Since $I$ is a prime ideal, 
it follows that $W$ belongs to $I$, hence there exist four polynomials $F_1,\ldots,F_4$ so that 
\eqn\pointA{
L_1F_1+L_2F_2+L_3F_3+L_4F_4=W.} 
Given this relation, we can find a factorization of $W$ of the form 
\eqn\pointB{ 
D = \sum_{a=1}^4 (L_a\pi_a + F_a\pib_a)}
where $(\pi_a,\pib_a)$ are free fermion operators generating a complex Clifford algebra. 
This was explained in section 2, equations \susyconstr\ and \bdrycharge.\ 
A special case of this construction is 
\eqn\specialcase{\eqalign{
&L_1= X_0-\eta X_1,\quad L_2=X_2,\quad L_3=X_3,\quad L_4=X_4\cr
& F_1={X_0^5+X_1^5\over X_0-\eta X_1},\quad F_2=X_2^4,\quad F_3=X_3^4,\quad F_4=X_4^4\cr}}
which corresponds to the object ${\oa^{(0)}_4}$ studied 
in detail before. Choosing $\IZ/5$ representations appropriately,  
\pointB\ yields a family of deformations of ${\oa^{(0)}_4}$.
We claim that the isomorphism classes of objects in this family are parameterized by points 
on the Fermat quintic. 
This is by no means obvious since a priori \pointB\ depends on the choice of a set of generators 
for the ideal $I$. 

In order to prove this claim, we have to rely on the results of \O.\ 
Suppose we are given a Landau-Ginzburg superpotential $W:\IC^{n+1}\ra C$ with an isolated critical 
point at the origin. Let $S_0$ denote the fiber of $W$ over $0\in \IC$. Then  
the main statement of \O\ is that the D-brane category 
$\CC_{W}$ is equivalent to the so-called category of the singularity 
$D_{Sg}(S_0)$. $D_{Sg}(S_0)$ is constructed by taking the quotient of the 
bounded derived category $D^b(S_0)$ by the full subcategory of perfect complexes. 
A perfect complex is a finite complex of locally free sheaves. If $S_0$ were nonsingular, 
the quotient would be empty, since in that case any object in $D^b(S_0)$ would have a 
locally free resolution. Therefore $D_{Sg}(S_0)$ depends only on the singular points
of $S_0$. The equivalence functor $\CC_{W} \ra D_{Sg}(S_0)$ associates to an object 
$\left(
\xymatrix{P_1 \ar@<1ex>[r]^{p_1}& P_0\ar@<1ex>[l]^{p_0}\\}\right)$
the one term complex defined by the cokernel of $p_1$ regarded as a coherent 
$\IC[X_0,\ldots, X_n]/W$-module.

A consequence of this result is that the isomorphism class of an object $\op$ 
is uniquely determined by 
the isomorphism class of the coherent $\IC[X_0,\ldots, X_n]/W$-module $\hbox{Coker}(p_1)$ 
modulo extensions by free $\IC[X_0,\ldots, X_n]/W$-modules. 
Using this result, it suffices to show that the cokernels $\hbox{Coker}(p_1)$ 
associated to the factorizations \pointB\ are parameterized by points on the Fermat quintic. 

Let $Q$ denote the cokernel associated to an arbitrary factorization \pointB\ and
let $R=\IC[X_0,\ldots,X_4]$.  
We claim that $Q$ is isomorphic to the 
$R/W$-module $R/I$ 
in $D_{Sg}(S_0)$. Therefore the deformations \pointB\ are parameterized up to 
isomorphisms 
by points on the Fermat quintic, as claimed above. 
Note that $R/I$ has an $R/W$-module structure because $W\in I$, according to \pointA.\ 

To determine $Q$, 
let us write down an explicit expression for the map $p_1$ associated to the 
factorization \pointB.\ 
Choosing an appropriate representation of the Clifford algebra, (or taking tensor products) we obtain 
\eqn\pointC{
p_1=\left[\matrix{ F_4 & 0 & 0 & 0 & L_3 & 0 & L_2 & F_1 \cr 
0 & F_4 & 0 & 0 & 0 & L_3 & L_1 & -F_2 \cr 
0 & 0 & F_4 & 0 & F_2 & F_1 & -F_3 & 0 \cr
0 & 0 & 0 & F_4 & L_1 & -L_2 & 0 & -F_3 \cr 
F_3 & 0 & L_2 & F_1 & -L_4 & 0 & 0 & 0 \cr
0 & F_3 & L_1 & -F_2 & 0 & -L_4 & 0 & 0 \cr
F_2 & F_1 & -L_3 & 0 & 0 & 0 & -L_4 & 0\cr
L_1 & -L_2 & 0 & -L_3 & 0 & 0 & 0 & -L_4\cr}\right]\quad .}
Let $U=\left[\matrix{U_1 & U_2 & \ldots & U_8}\right]^{tr}$ be 
an arbitrary element of $R^{\oplus 8}$. We want to determine the quotient module 
$Q=R^{\oplus 8} / (p_1(R^{\oplus 8}))$. Let $G=\left[\matrix{G_1 & G_2 & \ldots & G_8\cr}
\right]^{tr}$
denote an arbitrary element in $R^{\oplus 8}$. We have $G\sim 0$ if and only if 
$G=p_1U$ for some $U$. In particular this yields the relation 
\eqn\pointD{ 
G_8 = L_1U_1-L_2U_2-L_3U_4-L_4U_8}
which shows that the projection to the eighth factor induces a surjective $R/W$-module 
homomorphism 
\eqn\surjmapA{Q\ra R/I\ra 0\,.} 
The kernel of this map is an $R/W$-module $K$ isomorphic to the quotient $R^{\oplus 7}/(p_1(M))$.
Here $M\subset R^{\oplus 8}$ is the submodule of $R^{\oplus 8}$ which leaves $G_8$ invariant. 
One can easily check that $U\in M$ can be parameterized as follows 
\eqn\pointE{ \eqalign{ 
& U_1= V_{12}L_2 + V_{13}L_3 + V_{14} L_4 \cr 
& U_2=V_{12} L_1 + V_{23}L_3 + V_{24} L_4 \cr 
& U_4 = V_{13} L_1 -V_{23} L_2 + V_{44} L_4 \cr 
& U_8 = V_{14} L_1 - V_{24} L_2 - V_{44} L_3 \cr}}
where $V_{12},V_{13},\ldots V_{44}$ are arbitrary polynomials and $U_3,U_5,U_6,U_7$ are also arbitrary. 
Evaluating $p_1$ on elements $U$ of this form yields 
\eqn\pointF{ \eqalign{
p_1U = & [\matrix{V_7 L_2 + V_5 L_3 &  V_7 L_1 + V_6 L_3 & F_4V_3-F_3V_7+F_2V_5+F_1V_6 
& V_5L_1-V_6L_2\cr}\cr &\matrix{
V_3L_2-V_5L_4 & V_3L_1 - V_6L_4 & -V_3L_3-V_7L_4 & 0 \cr}]^{tr}\ \hbox{mod}\ W\cr}}
where 
\eqn\pointG{\eqalign{ 
& V_3 = U_3 -F_1 V_{23} -F_2 V_{13}+F_3 V_{12}\cr
& V_5 = U_5 -F_1V_{44}-F_3V_{14}+F_4V_{13} \cr
& V_6= U_6 +F_2V_{44}-F_3V_{24} +F_4V_{23} \cr
& V_7 = U_7 -F_1V_{24} -F_2V_{14} +F_4V_{12}\,.\cr}}
This shows that there is an exact sequence of $R/W$-modules of the form 
\eqn\surjmapB{ 
(R/W) ^{\oplus 4} {\buildrel f\over \ra} (R/W)^{\oplus 7} \ra K \ra 0}
where $f$ is defined by \pointF.\ 
The kernel of $f$ consists of elements $V = \left[\matrix{V_3 & V_5 & V_6 & V_7 \cr}\right]^{tr}$ 
of the form 
\eqn\pointH{ 
V_3=L_4V,\quad V_5=L_2V,\quad V_6=L_1V,\quad V_7=-L_3V,}
where $V$ is an arbitrary polynomial.  
Therefore $\hbox{Ker}(f)$ is isomorphic to $R/W$, and we conclude that $K$ has a finite free 
resolution 
\eqn\pointI{ 
 0\ra R/W \ra (R/W) ^{\oplus 4} {\buildrel f\over \ra} (R/W)^{\oplus 7} \ra K \ra 0}
as an $R/W$-module. 
This shows that $K$ is isomorphic to the trivial element in $D_{Sg}(S_0)$, hence 
$Q$ is indeed isomorphic to $R/I$ as claimed above. 

\appendix{A}{Morphism Spaces: Matrix to Polynomial Factorizations}

In this section, we study the space of morhisms between the fractional branes $\om^2$ and the 
rank one objects $\op_{\eta}$ in $\CC_{W}$ with $W=X_0^d+X_1^d$.
\ifig\matrixtopoly{The morphism complex for matrix to polynomial factorizations.}
{\epsfxsize5.0in\epsfbox{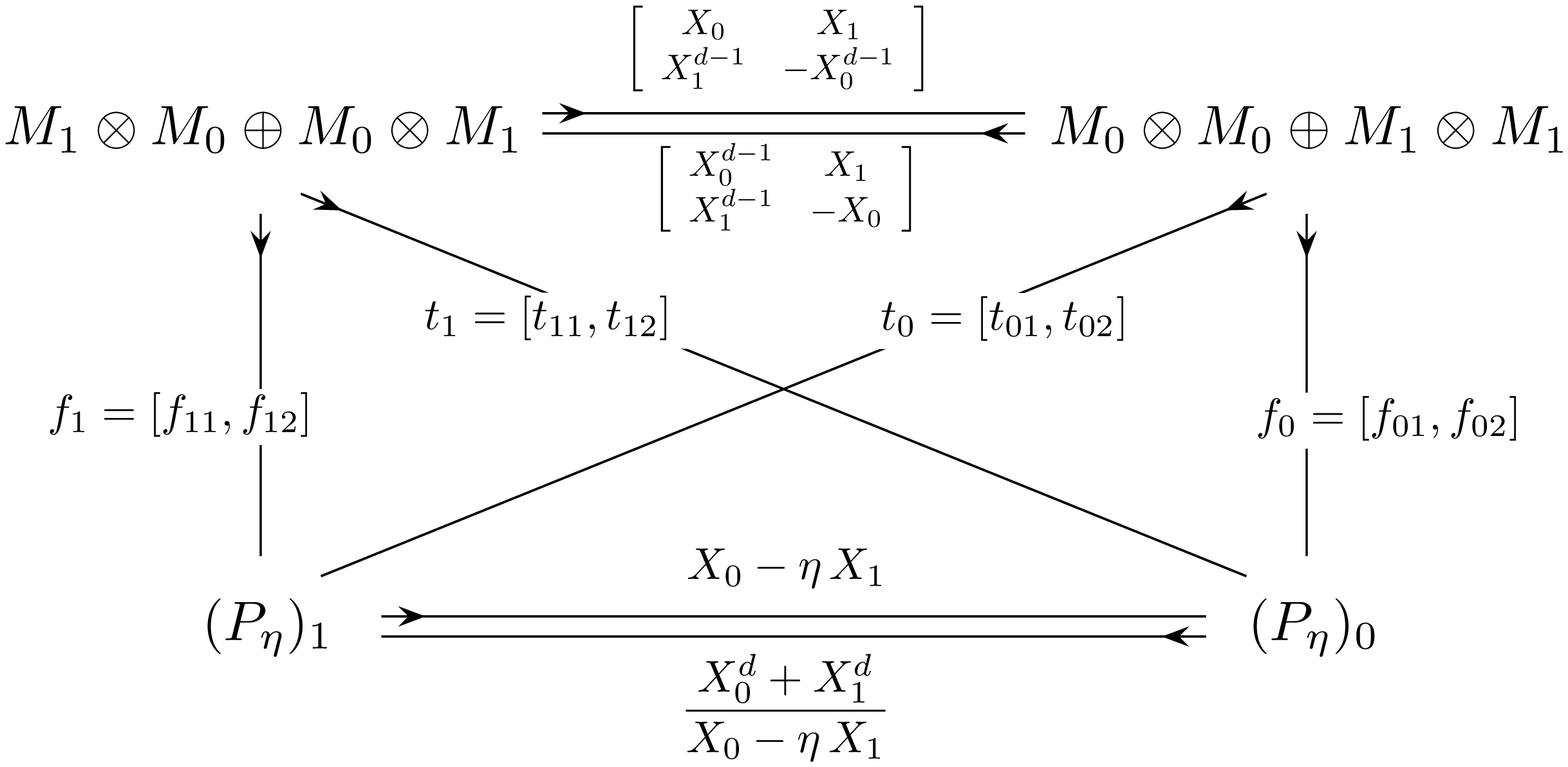}}
\noindent
For the bosonic morphisms, the conditions for $Q$-closedness takes the form 
\eqn\matQc{\eqalign{
(f_{01}\, X_0 + f_{02}\, X_1^{d-1})-(X_0-\eta X_1) f_{11} &= 0 \cr
(f_{01}\, X_1 - f_{02}\, X_0^{d-1})-(X_0- \eta X_1) f_{12} &= 0\,. 
}}
The $Q$-exactness condition for $f_{02}$
$$
f_{02}\sim 0 \quad \hbox{if} \quad f_{02} = (t_{11} X_1 - t_{12}\, X_0)+ t_{02}\, (X_0-\eta X_1)  
$$
allows one to set $f_{02}=1$. Substituting this into \matQc\ gives 
$$
[f_{11}, f_{12}]=\left[\eta^{d-1}X_1^{d-2}\,,\, -{X_0^d+X_1^d\over X_0-\eta X_1}
\right] \quad \hbox{and} \quad
\left[f_{01}, f_{02}\right]=\left[\eta^{d-1}X_1^{d-2}\,, \,1\right]\,.
$$
A similar computation for the fermionic morphisms gives 
$$
\left[t_{01}, t_{02}\right] = \left[-\eta\,{X_0^d+X_1^d\over X_0-\eta X_1}\,,\, 1\right] 
\quad \hbox{and} \quad 
\left[t_{11}\,,  t_{12}\right]=\left[\eta\,, 1\right]\,.
$$

\subsec{$\IZ_d$ Orbifold}

As before, imposing the equivariance conditions allows us to infer the 
morphisms between orbifolded objects. For instance,
$$
f_{02}: \left(M_1\otimes M_1\right)_{\mu-1}\longrightarrow \left(P_0\right)_{\mu'+1}
$$
is the identity operator therefore non-zero bosonic morphisms exist only for 
$\mu'-\mu=d-2$. 

Similarly for the fermions, since
$$
t_{02}: \left(M_1\otimes M_1\right)_{\mu-1}\longrightarrow \left(P_1\right)_{\mu'}
$$
is the identity operator, fermionic morphisms exist for $\mu'-\mu=d-1$. Thus, 
we infer that the intersection matrix from the matrix to the polynomial 
factorizations is
$$
\chi\left(\om_{11,\mu},\op_{\eta,\mu'}\right)=\left[G^{-2}-G^{-1}\right]_{\mu\mu'}
$$
where $G$ is the shift matrix.

\appendix{B}{Composition of morphisms for $W=X_0^5+X_1^5$}

As shown in \Tmatrices, the fermionic morphisms from $\overline{M}^2_1$ to $\overline{M}^2_0$ have the form
$$
T_0=\left(\matrix{
0 & 0 & 1 & 0 \cr
0 & 0 & 0 & X_0^3  \cr
-X_0^3 & 0 & 0 & 0  \cr
0 & -1 & 0 & 0 
}\right)\qquad\hbox{and}\qquad 
T_1=\left(\matrix{
0 & 0 & 0 & 1 \cr
0 & 0 & -X_1^3 & 0  \cr
0 & 1 & 0 & 0  \cr
-X_1^3 & 0 & 0 & 0 
}\right)\, .
$$
The morphisms from $\overline{M}^2$ to $\overline{P}_\eta$ have been constructed in Appendix A.
The fermionic morphism from $\overline{M}^2_0$ to $(\overline{P}_\eta)_4$ has the 
form\foot{For simplicity we suppress the subscripts $(\mu,\nu)$ used in section five.}
$$
t=\left(\matrix{
0 & 0 & t_{11} & t_{12} \cr
t_{01} & t_{02} & 0 & 0  
}\right) 
$$
with
$$
\left[t_{01}\,,\, t_{02}\right] = 
\left[-\eta \,(X_0^3 + \eta X_0^2 X_1 + \eta^2 X_0 X_1^2 + \eta^3 X_1^3)\,,\, 1\right] 
\quad \hbox{and} \quad 
\left[t_{11}\,,\,  t_{12}\right]=\left[\eta\,, 1\right]\, .
$$
The bosonic morphism from $\overline{M}^2_1$ to $(\overline{P}_\eta)_4$ has the form
$$
f=\left(\matrix{
 f_{01}& f_{02} & 0 & 0 \cr
0 & 0 & f_{11}& f_{12}  
}\right)
$$
with
$$
[f_{11}\, , f_{12}]=\left[\eta^{4}X_1^{3}\,, \,
-(X_0^3 + \eta X_0^2 X_1 + \eta^2 X_0 X_1^2 + \eta^3 X_1^3)\right] 
\quad \hbox{and} \quad
\left[f_{01}\,, f_{02}\right]=\left[\eta^{4}X_1^{3}\,, 1\right]\, .
$$
We want to show explicitly that the morphisms in the diagram of \EtoP\ compose as 
$$ 
t\,T_1J = \eta f J,\qquad 
t\,(\eta\, T_0 + T_1)J=0\,,
$$
where $J$ is introduced to account for the shift of the object $\overline{M}^2_1[1]\,$. Since here $J$ multiplies all expressions on the right, we can neglect it in the following computations. Using the expressions above we find
$$
t\, T_1 = \left(\matrix{
-X_1^3 & \eta & 0 & 0 \cr
0 & 0 & -X_1^3 &  -\eta\,\big(X_0^3 + \eta X_0^2 X_1 + \eta^2 X_0 X_1^2 + \eta^3 X_1^3 \big) 
}\right) = \eta\, f
$$
and
\eqn\Tont{\eqalign{
&t\,(\eta\, T_0 +T_1) =\cr -\eta^2&\left(\matrix{ X_0^3 -\eta^3 X_1^3 & 0 & 0 & 0 \cr
0 & 0 &  X_0\,(X_0^2+\eta X_0 X_1 +\eta^2 X_1^2) &  
 X_1\,\big(X_0^2+\eta X_0 X_1 +\eta^2 X_1^2\big) 
}\right)}
}
We need to show that the morphism that appears on the right hand side of this equality is zero in cohomology. A bosonic morphism $g$ is exact if
$$
g = D_{\overline{M}^2}\, s + s\, D_{\overline{P}_\eta} \,,
$$
with $s$ a fermionic morphism. The differentials $D_{\overline{M}^2}$ and 
$D_{\overline{P}_\eta}$ have the form
$$
 D_{\overline{M}^2} = \left(\matrix{
0 & 0 & X_0 & X_1 \cr
0 & 0 & X_1^4 & -X_0^4  \cr
X_0^4 & X_1 & 0 & 0  \cr
X_1^4 & -X_0 & 0 & 0 
}\right)\qquad\hbox{and}\qquad 
D_{\overline{P}_\eta} = \left(\matrix{
0 & X_0-\eta X_1 &   \cr
{X_0^5 + X_1^5}\over{X_0-\eta X_1} & 0 
}\right)\,.
$$
If we parameterize the morphisms $g$ and $s$ as
$$
g = \left(\matrix{
 g_{01}& g_{02} & 0 & 0 \cr
0 & 0 & g_{11}& g_{12}  
}\right)\qquad\hbox{and}\qquad 
s=\left(\matrix{
0 & 0 & s_{11} & s_{12} \cr
s_{01} & s_{02} & 0 & 0  
}\right)\,,
$$
the condition for $g$ to be exact gives the system of equations
$$\eqalign{
 g_{01}&= s_{11}\, X_0^4 + s_{12}\, X_1^4 + s_{01}\, (X_0-\eta X_1)\cr
g_{02} &= -  s_{12}\, X_0 +  s_{11}\, X_1 + s_{02}\, (X_0-\eta X_1)\cr
g_{11}&= s_{01}\, X_0 + s_{02}\, X_1^4 + s_{11}\, {X_0^5 + X_1^5\over X_0-\eta X_1}\cr
g_{12} &= -  s_{02}\, X_0^4 +  s_{01}\, X_1 + s_{12}\, {X_0^5 + X_1^5\over X_0-\eta X_1}
\,.}
$$
If we take $g$ to be \Tont, it is easy to see that these conditions can be satisfied by choosing $s_{01}=-\eta^2 \,(X_0^2+\eta X_0 X_1 +\eta^2 X_1^2)$ and $s_{02}=s_{11}=s_{12}=0\,$.

\listrefs
\end